\documentclass[%
 aip,
 amsmath,amssymb,
 reprint,%
]{revtex4-1}

\preprint{AIP/123-QED}

\usepackage{dcolumn}
\usepackage{bm}

\usepackage[utf8]{inputenc}
\usepackage[T1]{fontenc}
\usepackage{mathptmx}

\usepackage{color}
\usepackage[subpreambles=true]{standalone}
\usepackage[utf8]{inputenc}
\usepackage{import}
\usepackage{stackengine, graphicx}
\graphicspath{{../figs/}{figs/}}

\usepackage{pgfplots}
\usepackage{amsmath} 
\usepackage{amssymb}
\usepackage{mathtools}
\usepackage{enumitem}
\usepackage{booktabs}

\usepackage{multirow}

\usepackage{pifont}

\usepackage{flowchart}
\usepackage{tikz}
\usetikzlibrary{shapes,arrows,external}
\usetikzlibrary{decorations.pathreplacing,angles,quotes, calligraphy,matrix,positioning,calc}

\usepackage{floatrow}
\usepackage[caption=false, labelformat=simple]{subfig}

\usepackage{algorithm}
\usepackage{algpseudocode}
\usepackage{url}
\usepackage{flushend}

\setlist{nosep}

\usepackage{hyperref}

\def\equationautorefname~#1\null{Eq.(#1)\null}

\hypersetup{
pdfnewwindow=true, colorlinks=true,
linkcolor=blue, anchorcolor=blue,
citecolor=blue, filecolor=blue,
menucolor=blue, urlcolor=blue}

\usepackage{titlesec}

\newcommand{\be}{\begin{equation}}
\newcommand{\ee}{\end{equation}}

\newcommand{\bea}{\begin{eqnarray}}
\newcommand{\eea}{\end{eqnarray}}

\newcommand{\ba}{\begin{align}}
\newcommand{\ea}{\end{align}}

\newcommand{\bd}{\begin{description}}
\newcommand{\ed}{\end{description}}

\newcommand{\bi}{\begin{itemize}}
\newcommand{\ei}{\end{itemize}}

\newcommand{\benum}{\begin{enumerate}}
\newcommand{\eenum}{\end{enumerate}}

\newcommand\wordcount{
    \immediate\write18{texcount -sub=section \jobname.tex  | grep "Section" | sed -e 's/+.*//' | sed -n \thesection p > 'count.txt'}
(  2053
  249
  495
  179
words)}

\begin{document}

\title{Evaluation of synthetic and experimental training data in supervised machine learning applied to charge state detection of quantum dots}
%
\author{J. Darulov\'a}
 \address{Theoretische  Physik,  ETH  Zurich,  8093  Zurich,  Switzerland}

 \author{M. Troyer}
 \address{Microsoft Quantum, Redmond, Washington 98052, USA}

 \author{M.C. Cassidy}
 \address{ Microsoft  Quantum,  The  University  of Sydney, Sydney, NSW 2006, Australia}

\date{\today}

\begin{abstract}

Automated tuning of gate-defined quantum dots is a requirement for large scale semiconductor
based qubit initialisation. An essential step of these tuning procedures is charge state
detection based on charge stability diagrams. Using supervised machine learning to perform this
task requires a large dataset for models to train on. In order to avoid hand labelling experimental
data, synthetic data has been explored as an alternative. While providing a significant increase in the size of the training dataset compared to using experimental data, using synthetic data means that classifiers are trained on data sourced from a different
distribution than the experimental data that is part of the tuning process. Here we
evaluate the prediction accuracy of a range of machine learning models trained on simulated and experimental
data and their ability to generalise to experimental charge stability diagrams in two dimensional electron gas and nanowire devices. We find that classifiers perform best on either purely experimental or a combination of synthetic and experimental training data, and that adding common experimental noise signatures to the synthetic data does not dramatically improve the classification accuracy. These results suggest that experimental training data as well as realistic quantum dot simulations and noise models are  essential in charge state detection using supervised machine learning.
\end{abstract}
\maketitle


A commercially viable quantum computer will require manufacturing of a large number of qubit devices, as well as autonomous procedures for qubit initialisation and tuning.
For semiconductor based qubits, such as charge \citep{PhysRevLett.105.246804, PhysRevLett.95.090502, Yang:2019wu}, spin \citep{PhysRevA.57.120, RevModPhys.79.1217, Petta2180, Veldhorst:tr} and topological qubits 
\citep{Kitaev:2001gb, Karzig:2017if, Alicea:2011fe}, this implies the initialisation of quantum dots of known charge state. These quantum dots are formed by choosing appropriate voltages on electrostatic gates, depleting the electron gas underneath and thus isolating electrons from surrounding charge carriers. 
Although the formation of these dots by human tuning is now routine in a wide range of systems, challenges still exist in automating this procedure.
Materials defects and fabrication variances result in non-uniform device performance, making tune-up procedures and operating gate voltages unique between nominally identical devices.

With machine learning and artificial intelligence succeeding in an increasing range of sophisticated tasks, it is worth investigating their capabilities in automating the task of defining quantum dots.
Several efforts in automating double quantum dot tune-up have been made using either supervised deep learning \cite{Kalantre2019, Zwolak2020, zwolak_qlite, Durrer2020}, unsupervised statistical methods \cite{Lennon:2019uq, Moon2020} or deterministic algorithms \cite{Darulova2020, Baart_automation, Lapointe-Major2019}. For these approaches to be useful for large scale tune-up, the tuning outcome  needs to be determined reliably despite noise and without human intervention. This is achieved by verifying the charge state of the qubit device based on a measured charge stability diagram.
Using supervised learning to perform this task requires a significant amount of labelled data, each with an attributed label indicating its charge state.
The process of measuring and labelling experimental data is slow, making synthetic data a way to increase the efficiency of this training process. The success of this technique has only been demonstrated on classifying further synthetic data, and so the usefulness of this approach has not been determined in classifying experimental data from real devices \cite{Kalantre2019, Zwolak2020, zwolak_qlite}.

Here we evaluate the ability of supervised machine learning models trained on synthetic data to determine the charge state of experimental charge stability diagrams, and compare to ones trained on data from real devices. Two convolutional neural network architectures and six parametric binary classifiers are trained to distinguish single versus double quantum dots when trained on purely synthetic data, a combination of synthetic and experimental data or experimental data only.  We also investigate how adding noise to noiseless synthetic data affects the classification accuracy.
 

Quantum dots are formed by applying voltages to electrostatic gates fabricated on top of a semiconductor structure, which creates  potential wells isolating charges in regions with length scales on the order of the Fermi wavelength. 
One or two regions of charges can be formed, resulting in a single or double quantum dot. To determine the regime, i.e. single versus double, two gate voltages are stepped over while the current through the device is measured, resulting in a so-called charge stability diagram. A single dot features sharp, diagonal lines, while a double dot shows either triple points with no charge transition lines between them, or a honey comb pattern with transition lines connecting bright spots corresponding to triple points. It has been shown that  voltage combinations not resulting in the formation of any dots can be excluded through simple gate characterisation steps\cite{Darulova2020}. Machine learning techniques are therefore only required to distinguish between single and double dots of different qualities to complete the tuning process.

In this work, we assess the accuracy of convolutional neural networks and binary classifiers trained on synthetic or experimental data. Convolutional neural networks trained and benchmarked on synthetic data have previously shown high classification accuracy \citep{Kalantre2019, zwolak_qlite, Zwolak2020}. We compare classification accuracies of both neural network architectures used, with one  \cite{zwolak_qlite} or two convolutional layers \cite{Kalantre2019, Zwolak2020} respectively. As the shallower network reaches lower accuracies, the main analysis is performed on a network consisting of two convolutional layers and four denser layers. The network's architecture is summarised in \autoref{tab:cnn}.

The binary classifiers we compare are logistic regression, multi-layer perceptron,  decision tree, random forest, K nearest neighbours and support vector machine, and  were selected based on the accuracy comparison in Ref. \cite{Darulova2020}.
All classifiers are trained and tested on the same data combinations. Binary classifiers are trained and tested on transport measurements and their frequency spectrum extracted using a Fourier transform \cite{Darulova2020}. Neural networks are trained and tested transport measurements only. Our techniques also apply to other types of measurements such as charge sensing and radio-frequency reflectometry.

\begin{figure}[!t]
    \begin{tabular}{c}
   \hspace{2em}  Double dot \hspace{6.5em} Single dot \hspace{1em} \\[-0.5em]
        \sidesubfloat[]{%
            \vspace{-2in}
      \hspace{-0.06in}%
            \includegraphics[width=0.44\columnwidth]{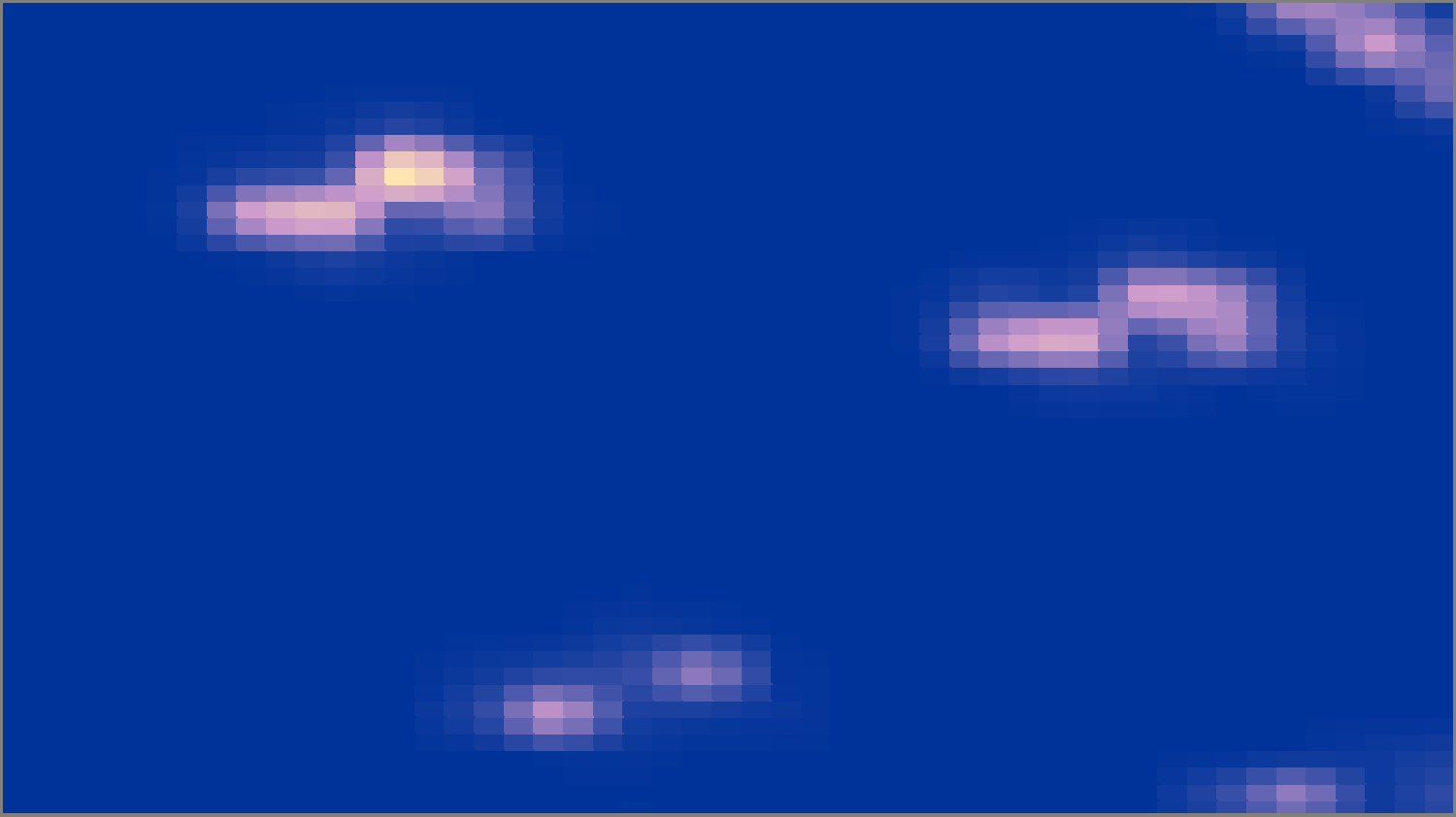}
            \includegraphics[width=0.44\columnwidth]{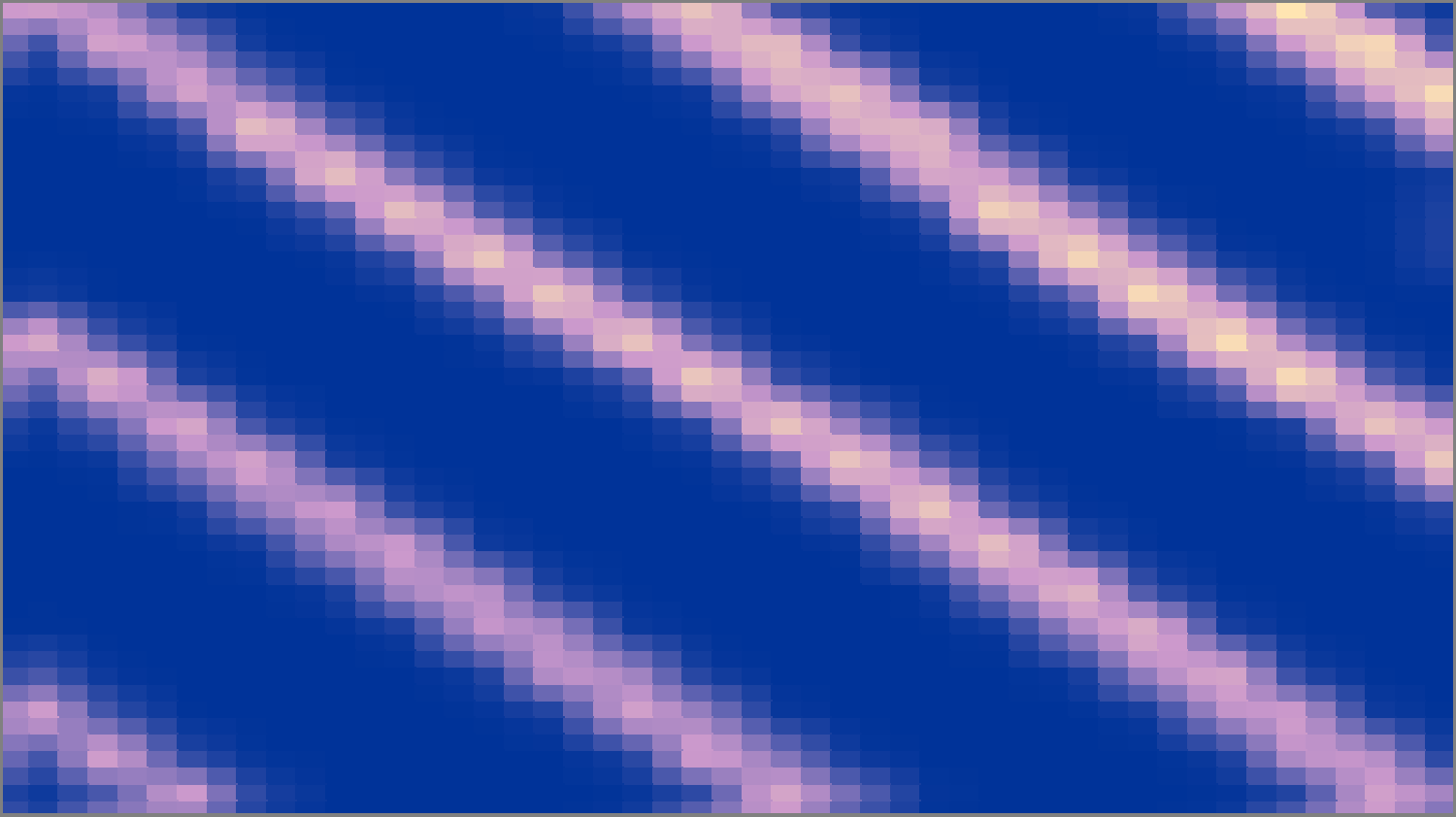} \label{fig:qflow_dd}
        }  \\
        \sidesubfloat[]{%
        \vspace{-2in}
      \hspace{-0.06in}%
            \includegraphics[width=0.44\columnwidth]{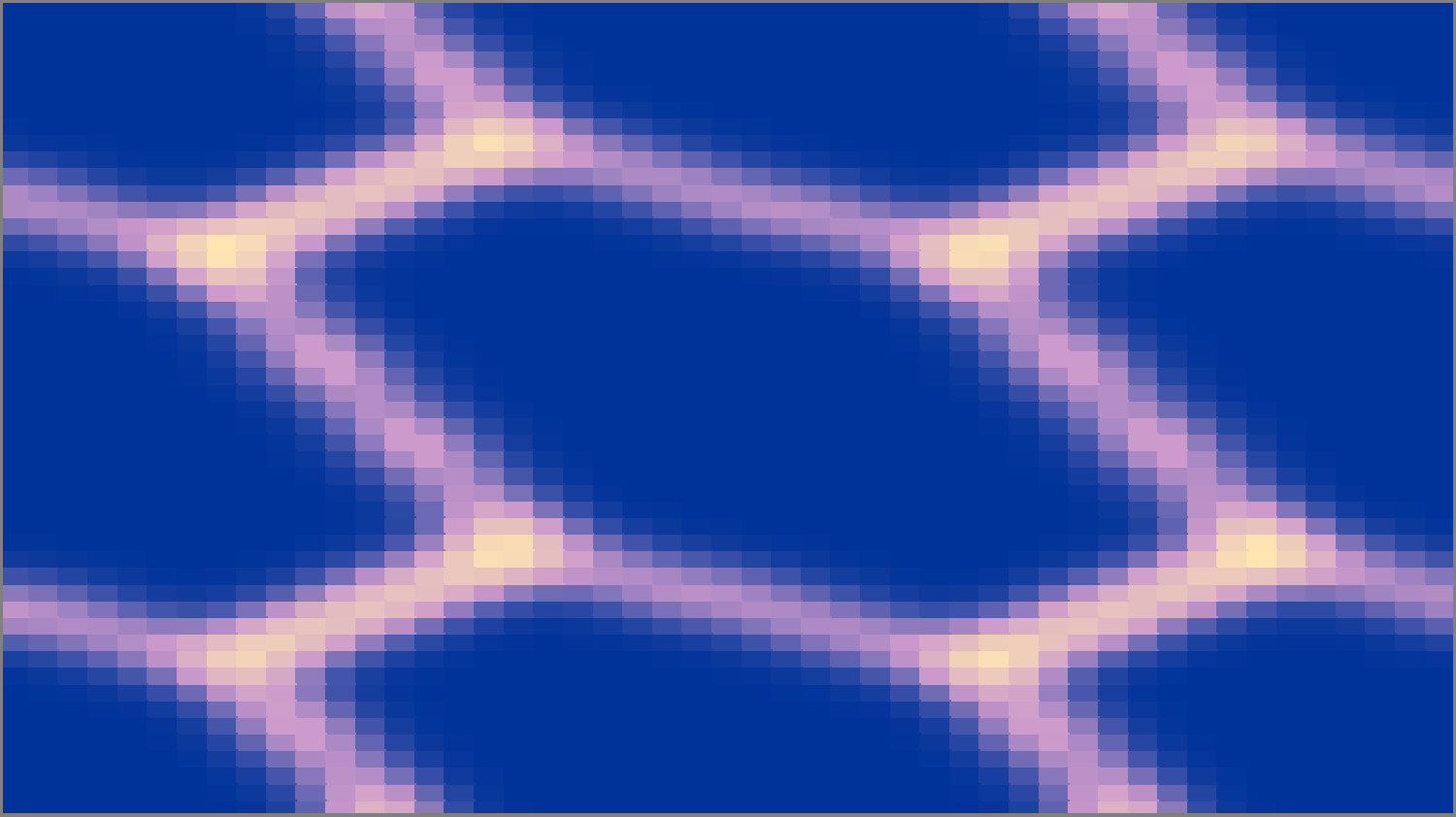}
            \includegraphics[width=0.44\columnwidth]{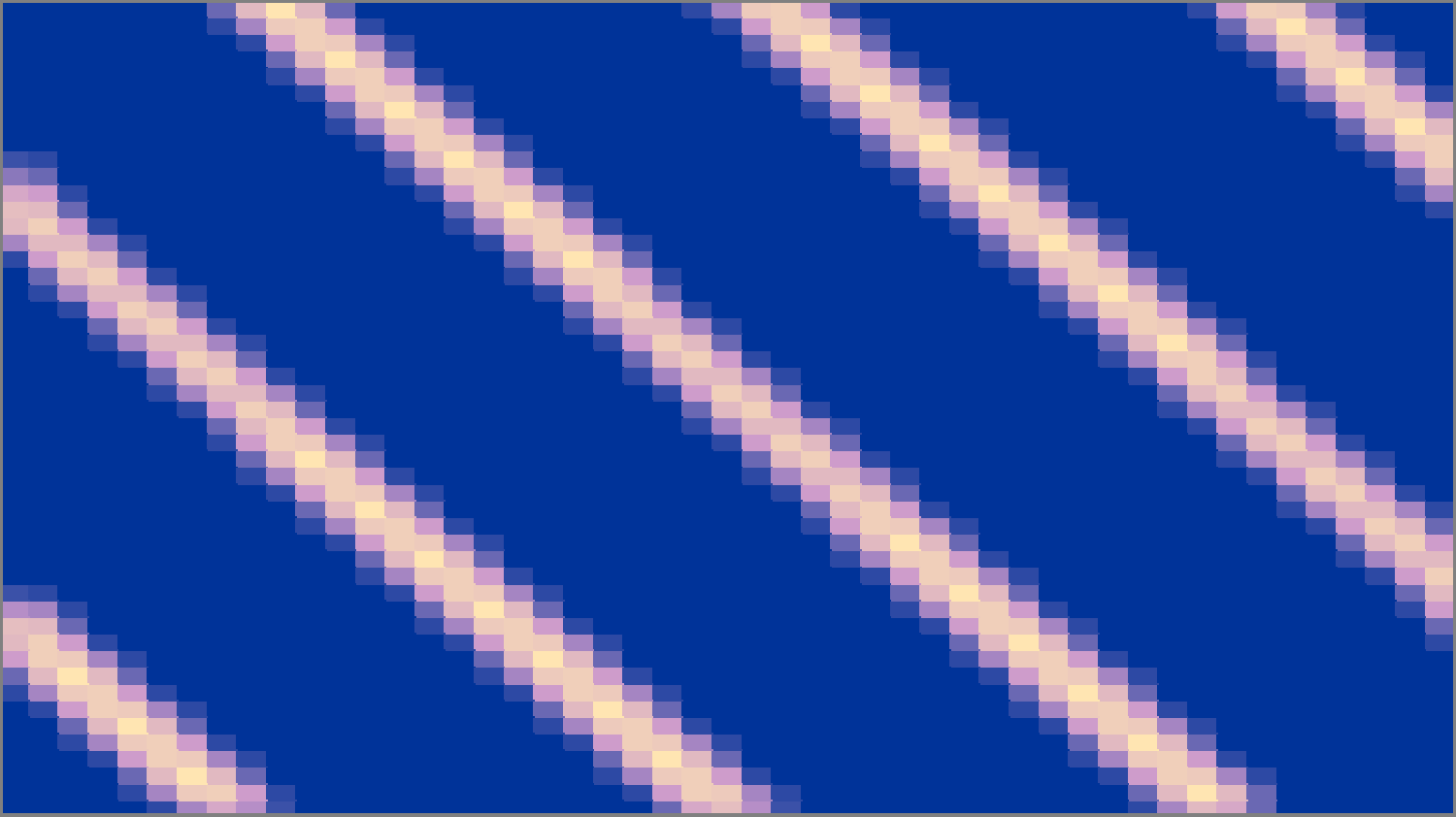} \label{fig:cm_dd}
        } 
    \end{tabular}
    \caption{Examples of noiseless synthetic charge stability diagrams. Left and right column show double and single dot regimes respectively.
            (a) Typical diagrams of the QFlow-lite dataset, simulates using the Thomas-Fermi approximation.%
            (b) Typical diagrams generated using the capacitance model. %
    }
  \label{fig:syn_examples}
\end{figure}

%

Our synthetic dataset of simulated single and double dot charge stability diagrams is based on data generated by a capacitance model \cite{VanderWiel2003} and the Qflow-lite dataset \cite{zwolak_qlite}, which uses the Thomas-Fermi approximation. Examples of both dot regimes generated by these models are shown in \autoref{fig:syn_examples}.
Details about the data generation and post-processing steps can be found in \autoref{ax:syn_data}.

\begin{figure}[!t]
    \begin{tabular}{c}
        \sidesubfloat[]{%
        \vspace{-2in}
        \hspace{-0.06in}%
            \includegraphics[width=0.4\columnwidth]{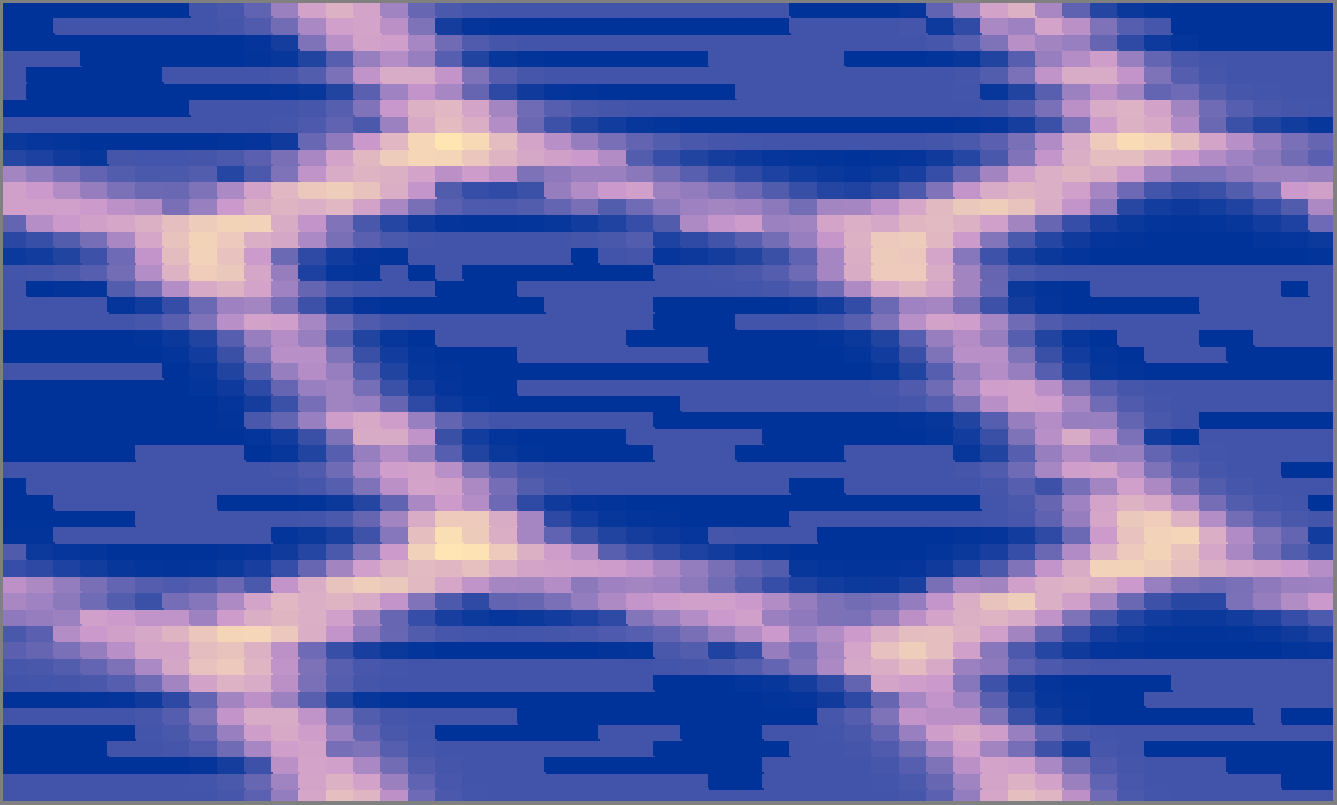}
        } 
        \sidesubfloat[]{%
        \vspace{-2in}
      \hspace{-0.06in}%
            \includegraphics[width=0.4\columnwidth]{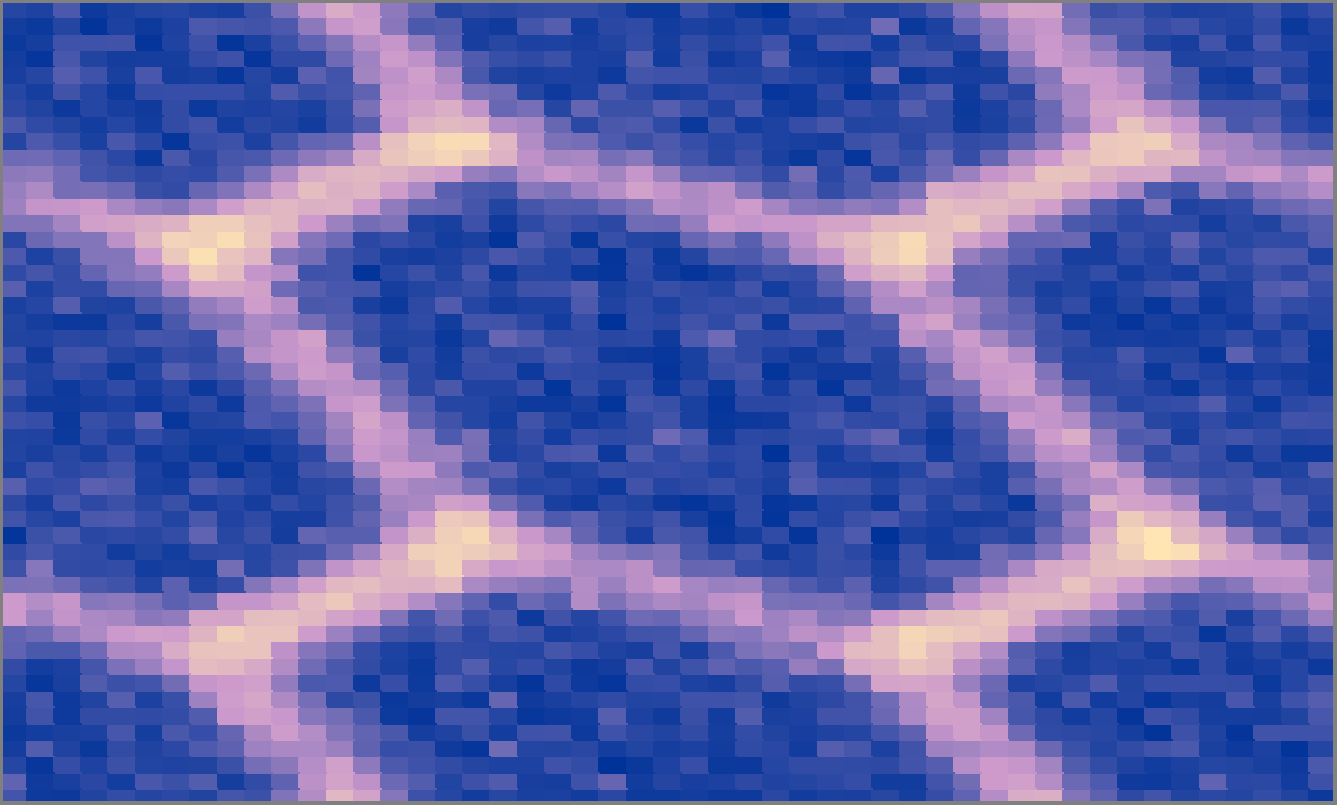}
        }
        \\ 
         \sidesubfloat[]{%
            \vspace{-2in}
      \hspace{-0.06in}%
            \includegraphics[width=0.4\columnwidth]{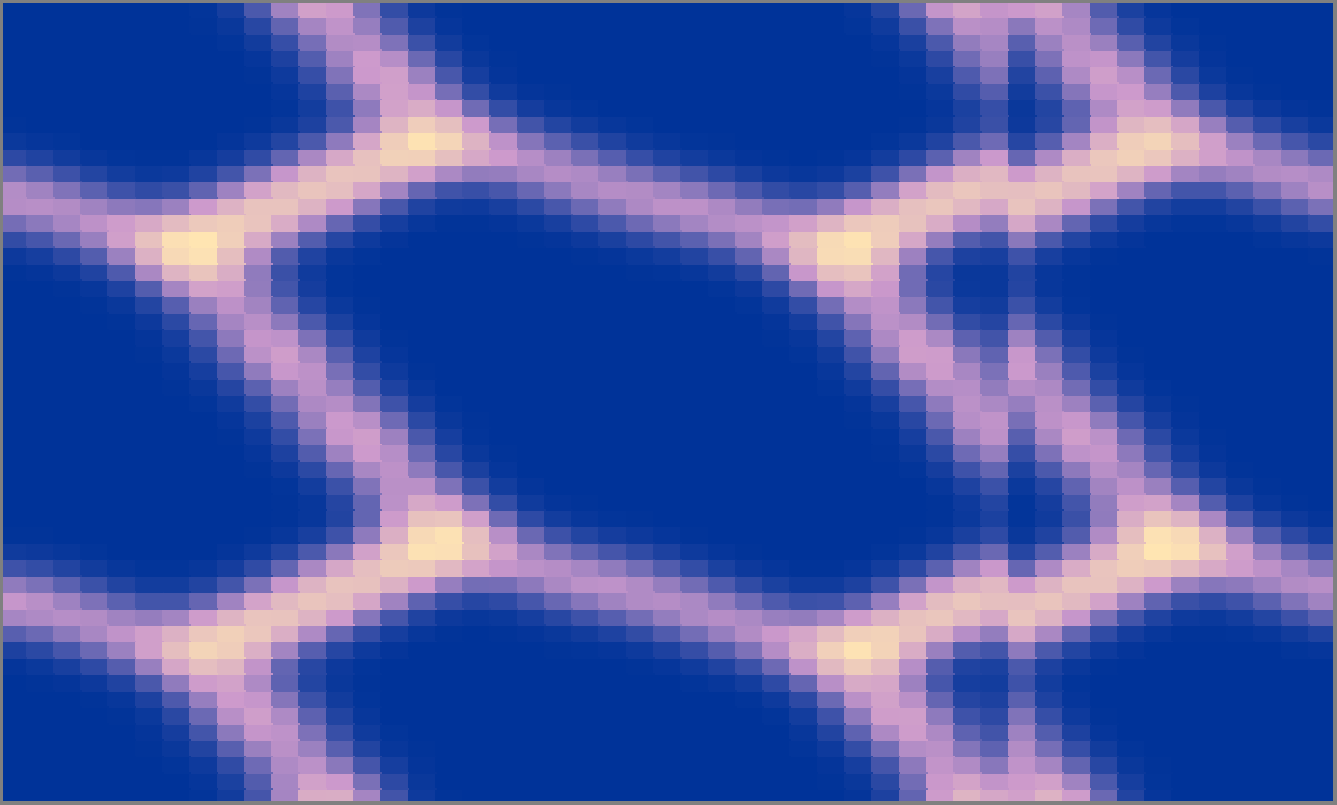}
        }
        \sidesubfloat[]{%
            \vspace{-2in}
      \hspace{-0.06in}%
            \includegraphics[width=0.4\columnwidth]{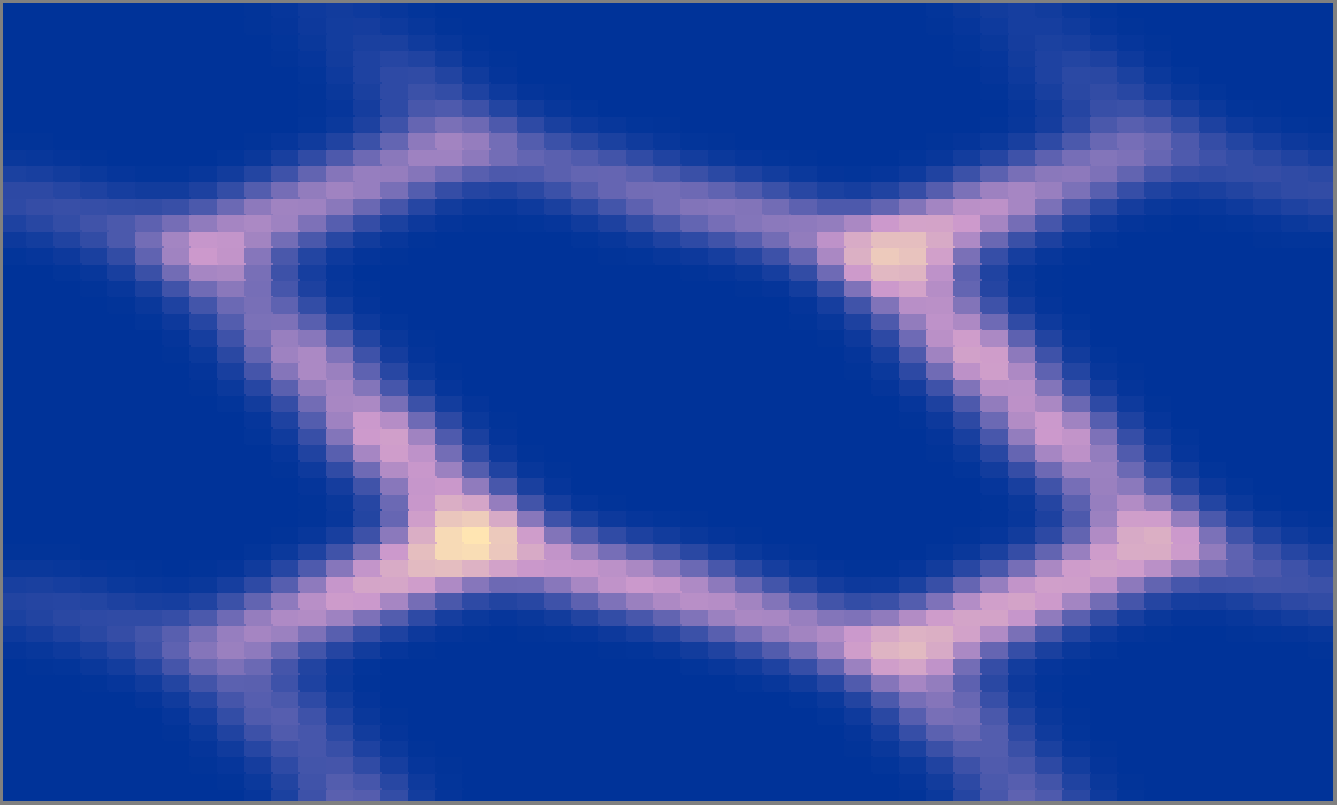} 
        }\\
        \sidesubfloat[]{%
            \vspace{-2in}
      \hspace{-0.06in}%
            \includegraphics[width=0.4\columnwidth]{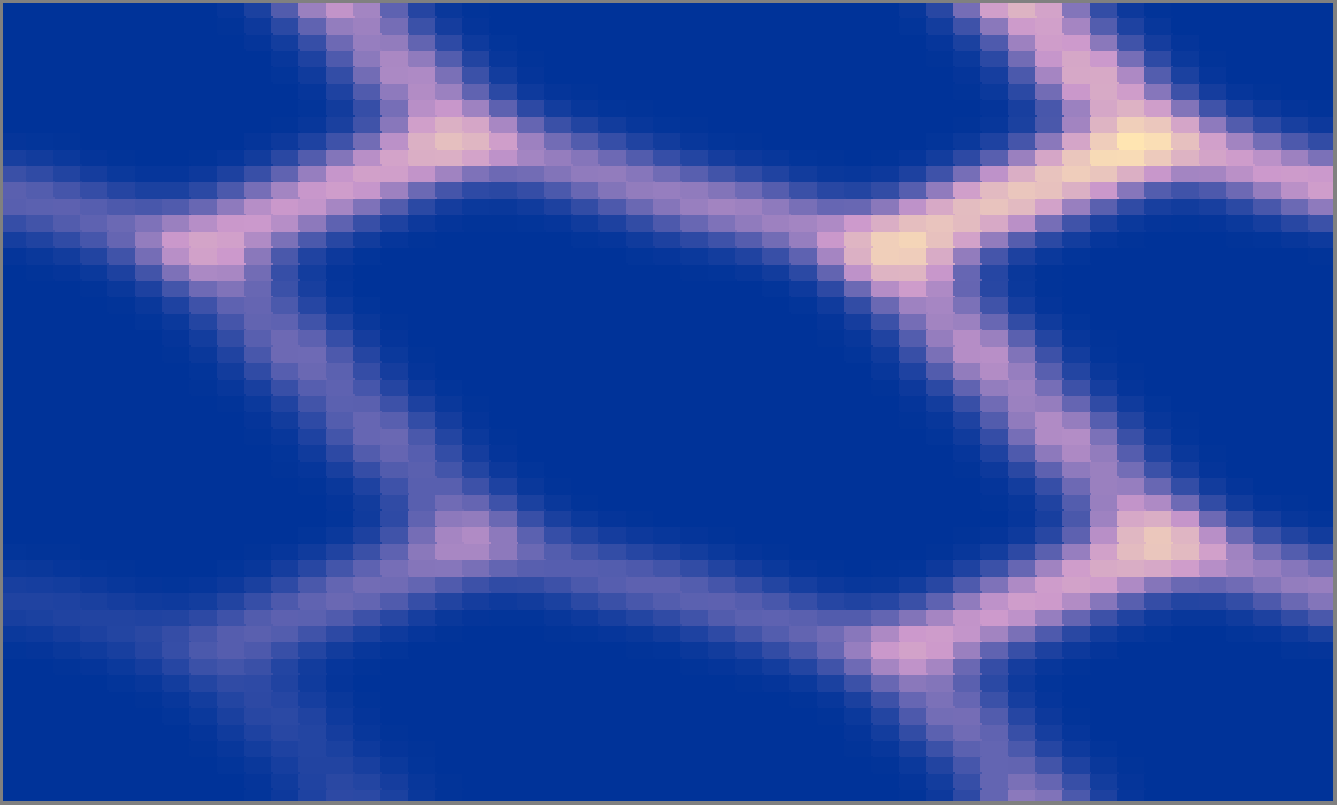}
        }
        \sidesubfloat[]{%
            \vspace{-2in}
      \hspace{-0.06in}%
            \includegraphics[width=0.4\columnwidth]{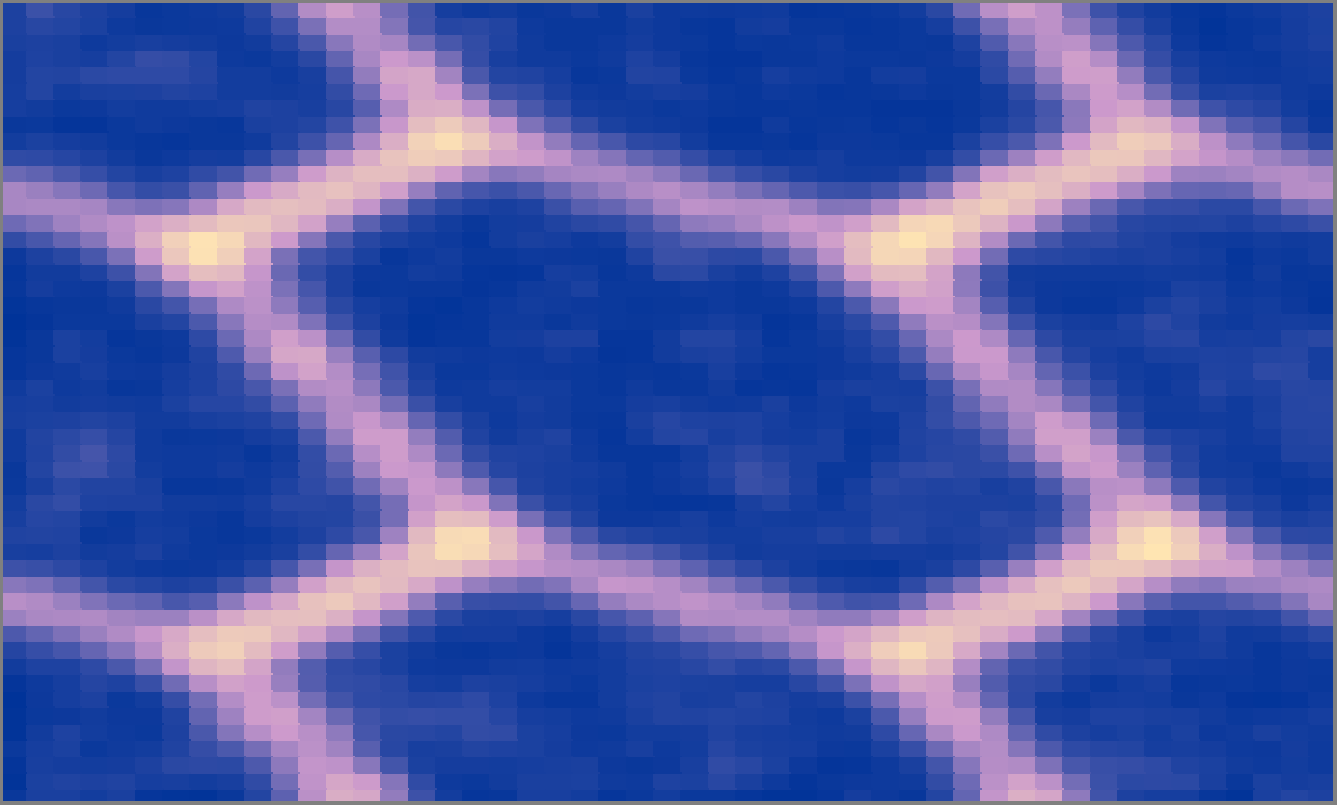} 
        }
    \end{tabular}
    \caption{Examples of one synthetic dataset with added noise. Random current maps were generated for each noise type independently, rescaled with varying amplitudes and then added to noiseless synthetic data. 
            (a) Random telegraph noise. %
            (b) White noise. %
            (c) Charge jumps. %
            (d) Random current modulation.
            (e) Pinchoff current modulation.
            (f) $1/f$ noise.
        }
     \label{fig:noise_examples}
\end{figure}

 We implement five noise models typically encountered in experiments that are added to the noiseless synthetic data. We refer to this dataset as noisy synthetic data set. These noise types are white noise, random telegraph noise, $1/f$ noise, charge fluctuations on gates, low frequency current modulations and pinch-off current modulation. An example of each is shown in \autoref{fig:noise_examples}. 
White noise typically arises due to thermal fluctuations, while $1/f$ noise and charge fluctuations on gates are two types of random fluctuations due to defects in the semiconductor. Random telegraph noise on the other hand is a low-frequency modulation of current caused by the spontaneous capture and emission of charge carriers. Low frequency current modulations and pinch-off current modulation are consequences of the electron gas being depleted for decreasing gate voltages. 
Details of their implementation as well as additional examples can be found in \autoref{ax:syn_noise}.

\begin{figure*}[!t]
    \begin{tabular}{c}
       \hspace{2em}  Double dot \hspace{19em} Single dot \hspace{1em} \\[-0.5em]
    \sidesubfloat[]{%
        \label{fig:clean_double}
        \vspace{-2in}
        \hspace{-0.1in}%
                                \noindent\stackinset{l}{1pt}{b}{0pt}{ 
                      {\setlength{\fboxsep}{1pt}\colorbox{white}{\includegraphics[width=0.07\columnwidth]{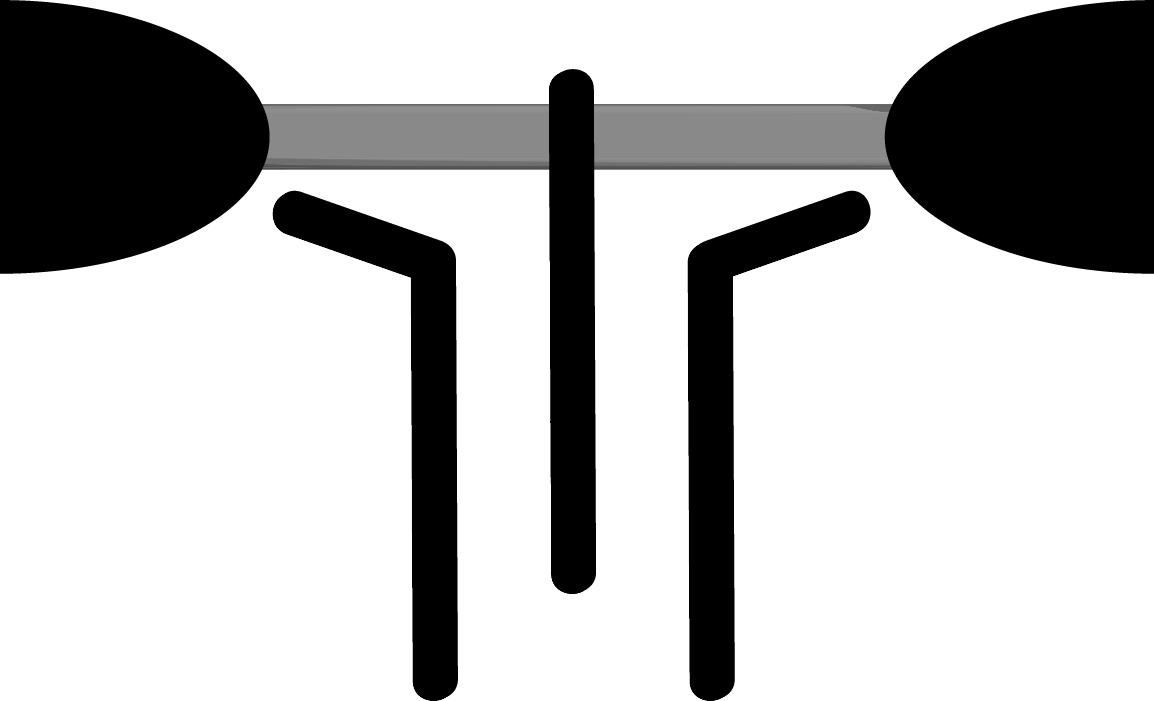}}}}
                        {     \includegraphics[width=0.2\columnwidth]{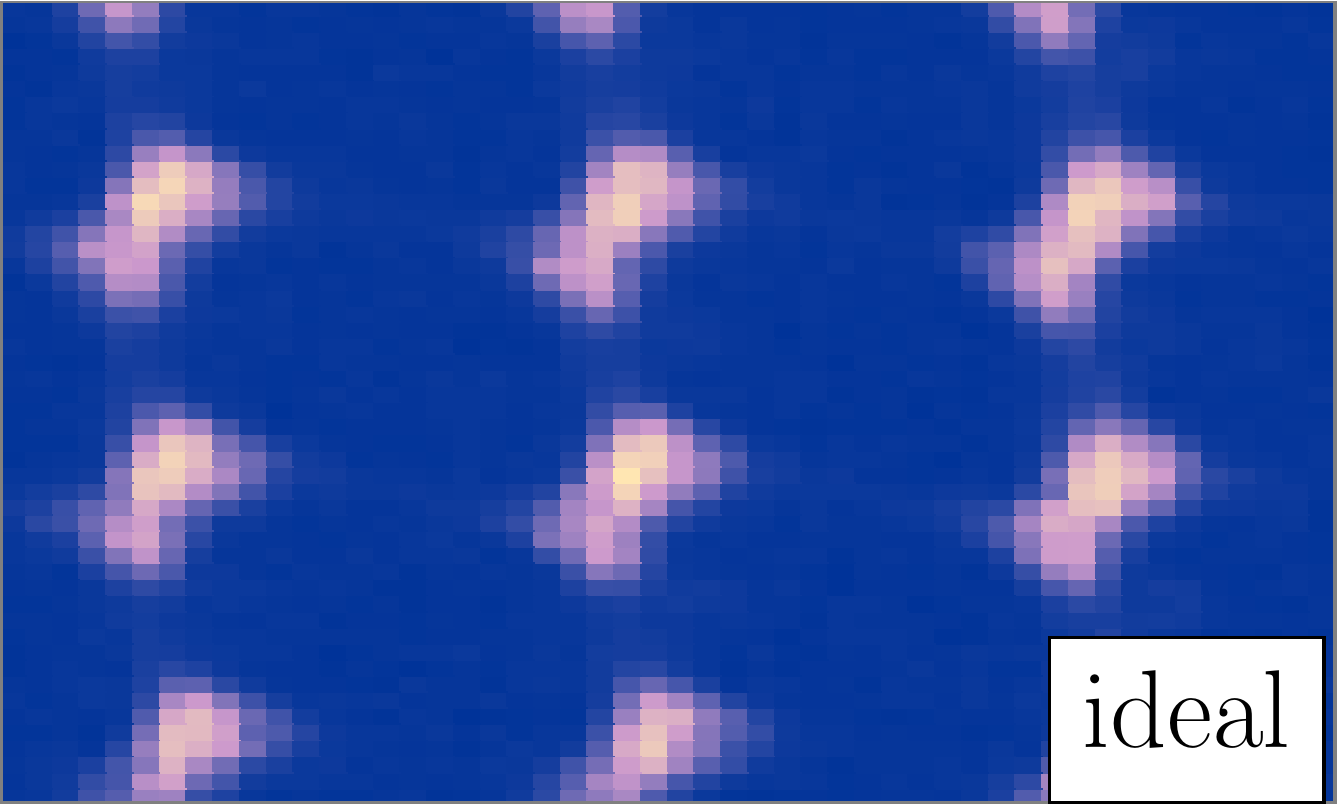}}
	   \includegraphics[width=0.2\columnwidth]{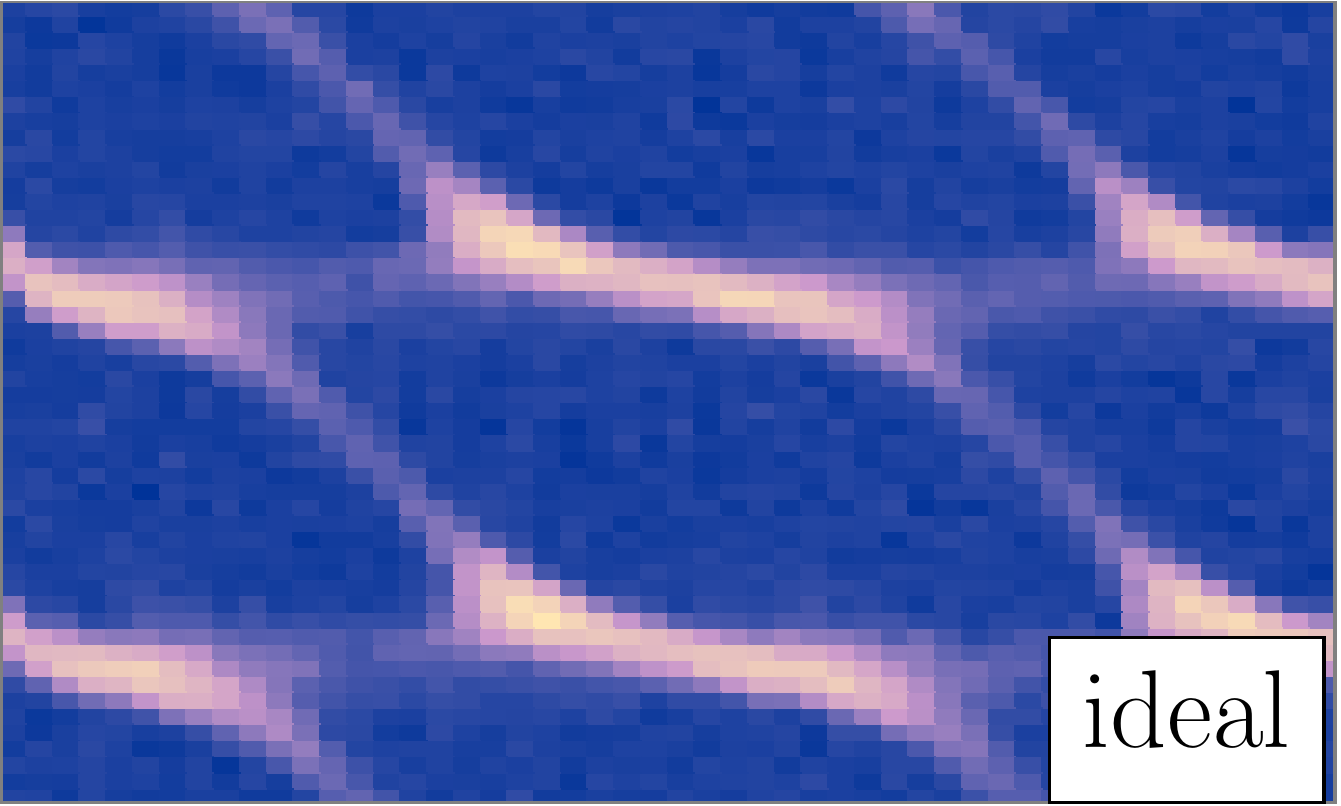}
        }  
        \sidesubfloat[]{%
        \label{fig:clean_single}
            \vspace{-2in}
            \hspace{-0.1in}%
                                            \noindent\stackinset{l}{1pt}{b}{0pt}{ 
                      {\setlength{\fboxsep}{1pt}\colorbox{white}{\includegraphics[width=0.045\columnwidth]{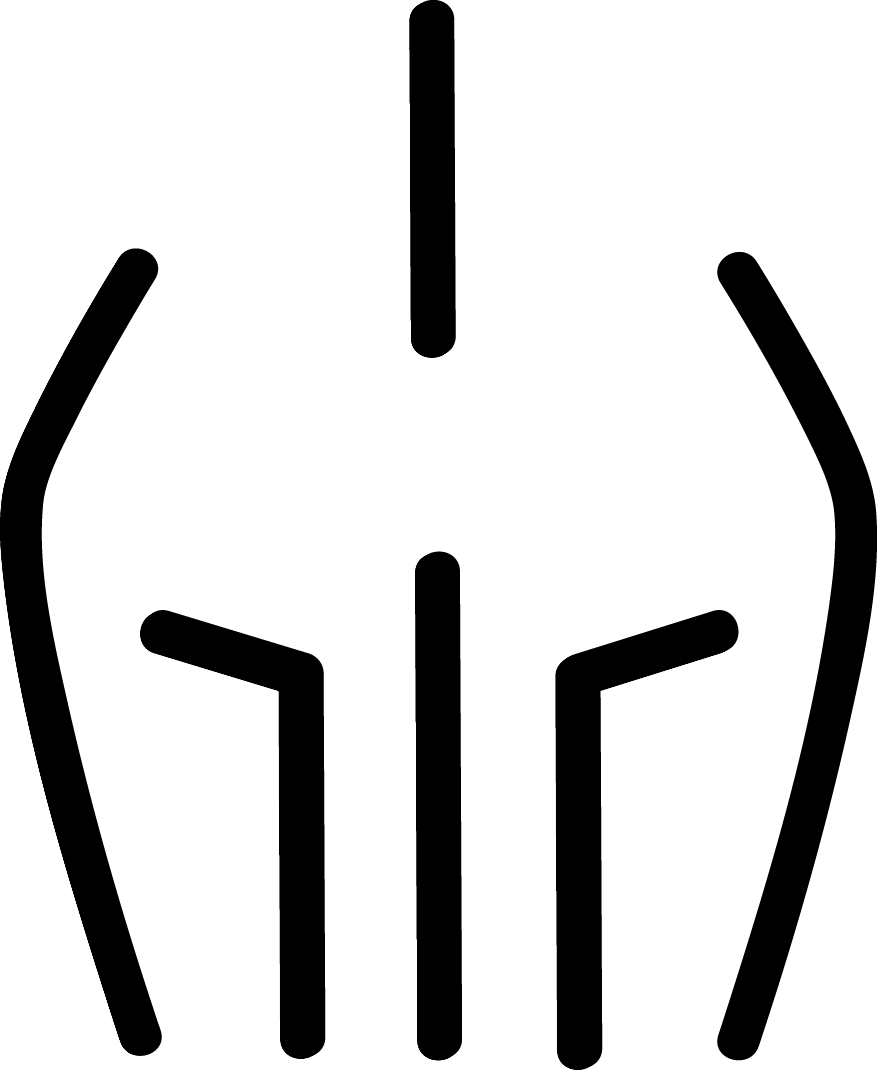}}}}
                        {                \includegraphics[width=0.2\columnwidth]{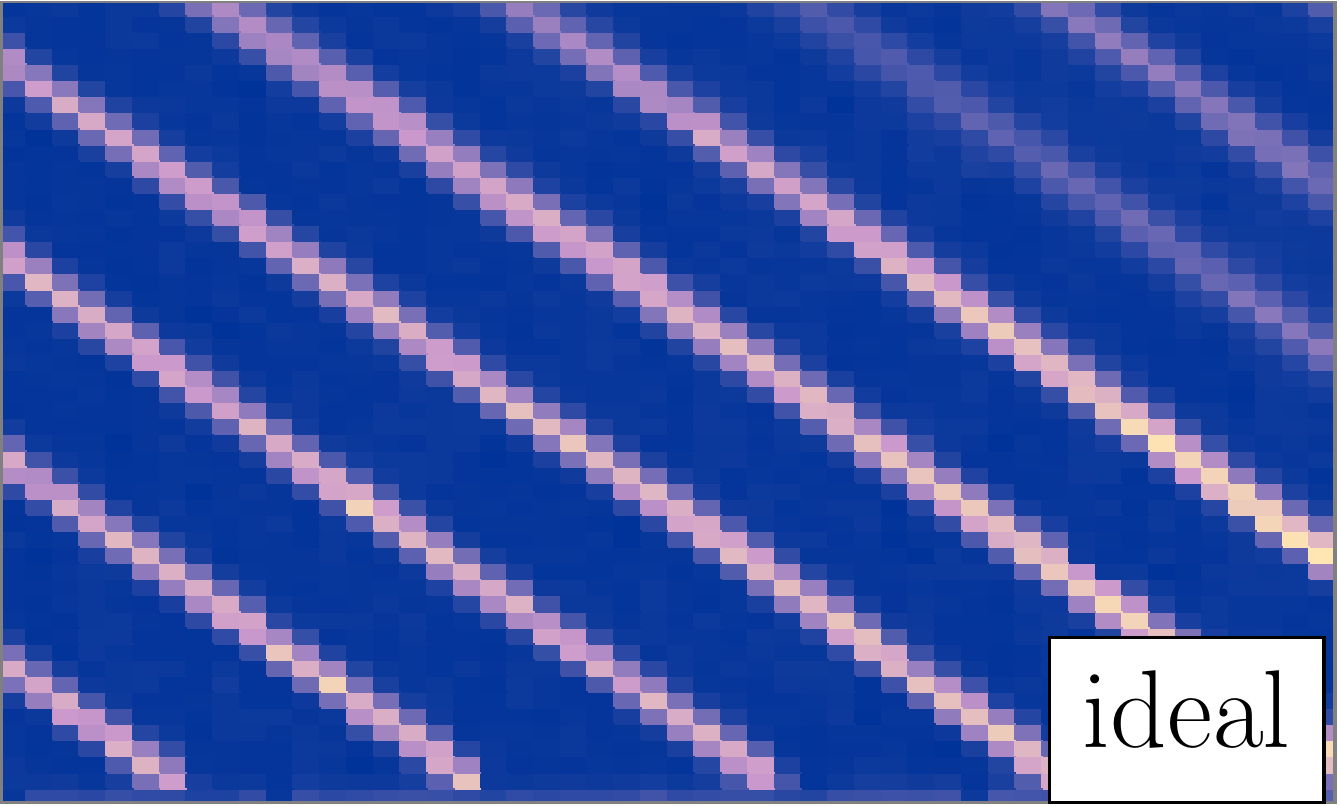} }
            \includegraphics[width=0.2\columnwidth]{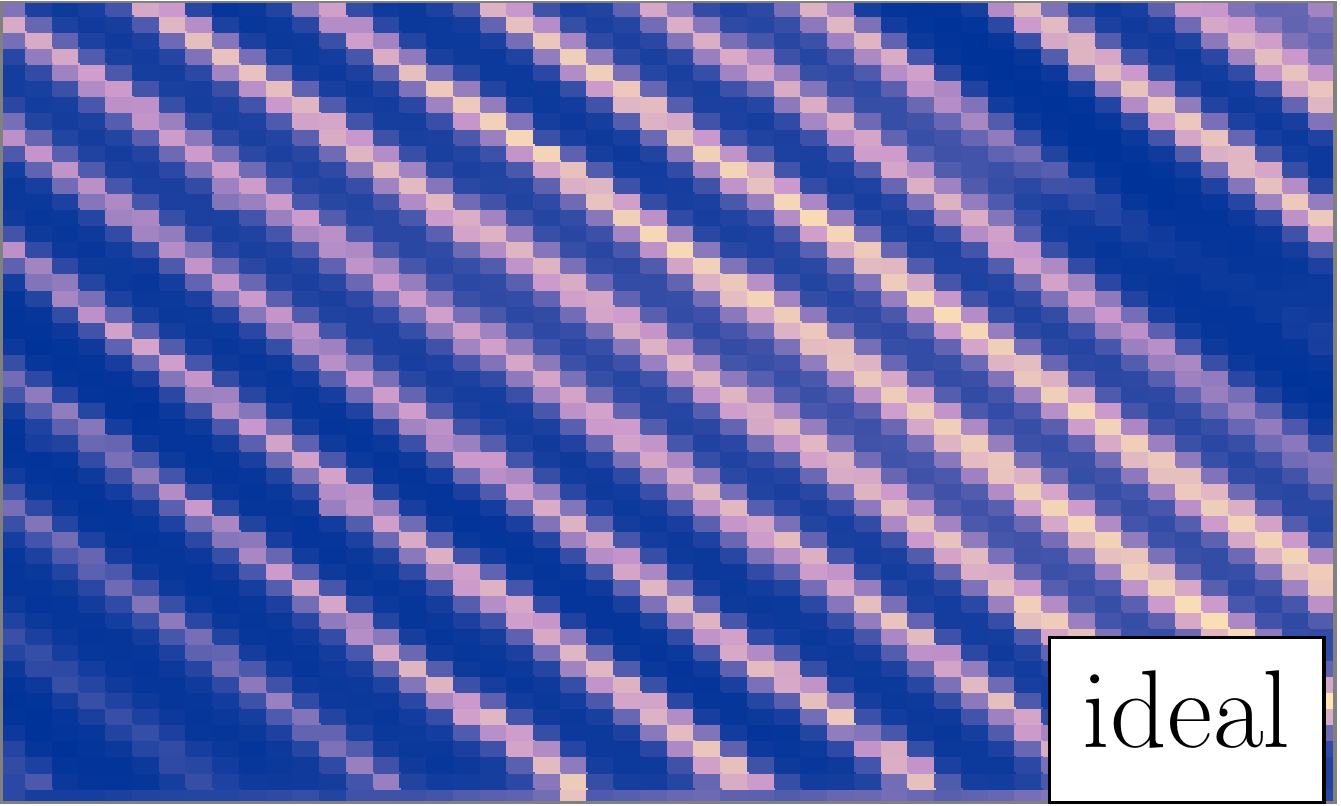}
        }
        \\
        \sidesubfloat[]{%
        \label{fig:good_double}
        \vspace{-2in}
        \hspace{-0.1in}%
                                                    \noindent\stackinset{l}{1pt}{b}{0pt}{ 
                      {\setlength{\fboxsep}{1pt}\colorbox{white}{\includegraphics[width=0.045\columnwidth]{cartoons-37}}}}
                        {                          \includegraphics[width=0.2\columnwidth]{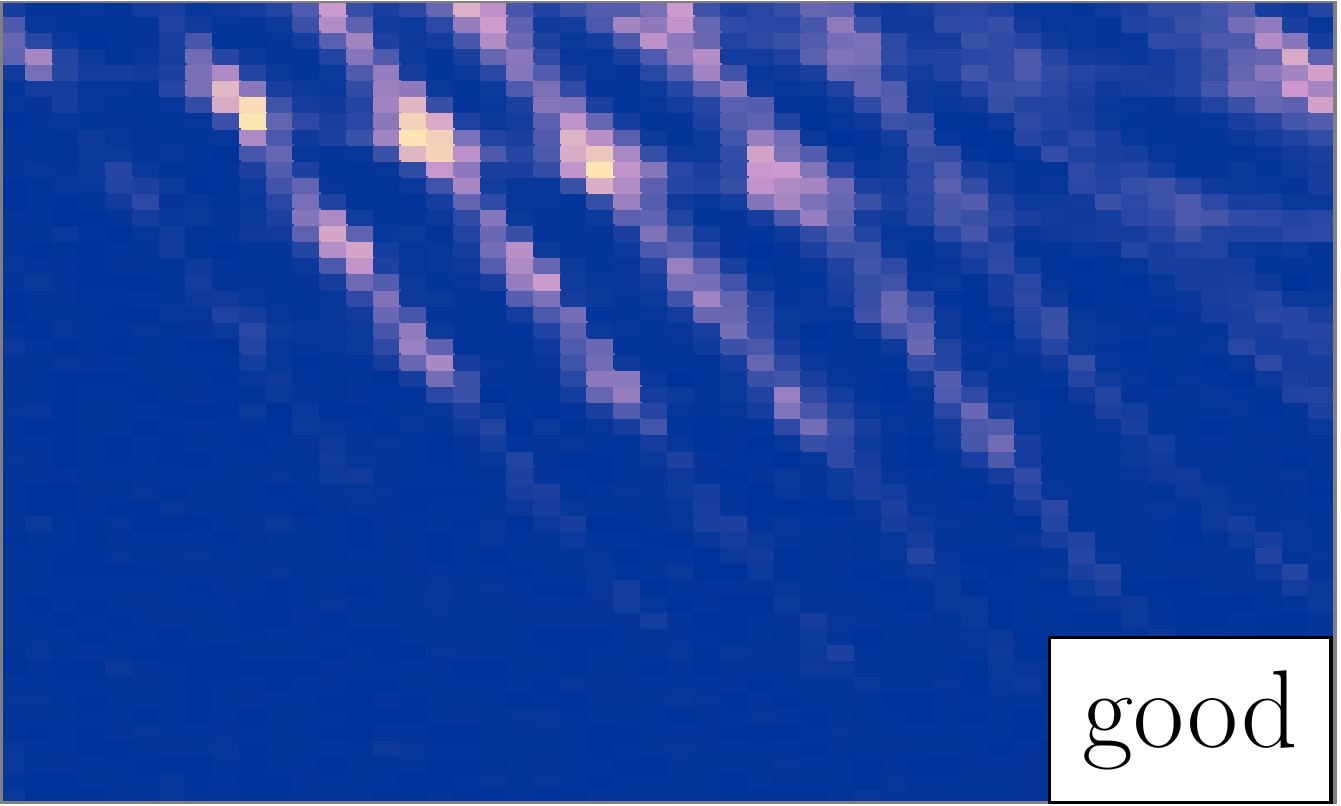}}
            \includegraphics[width=0.2\columnwidth]{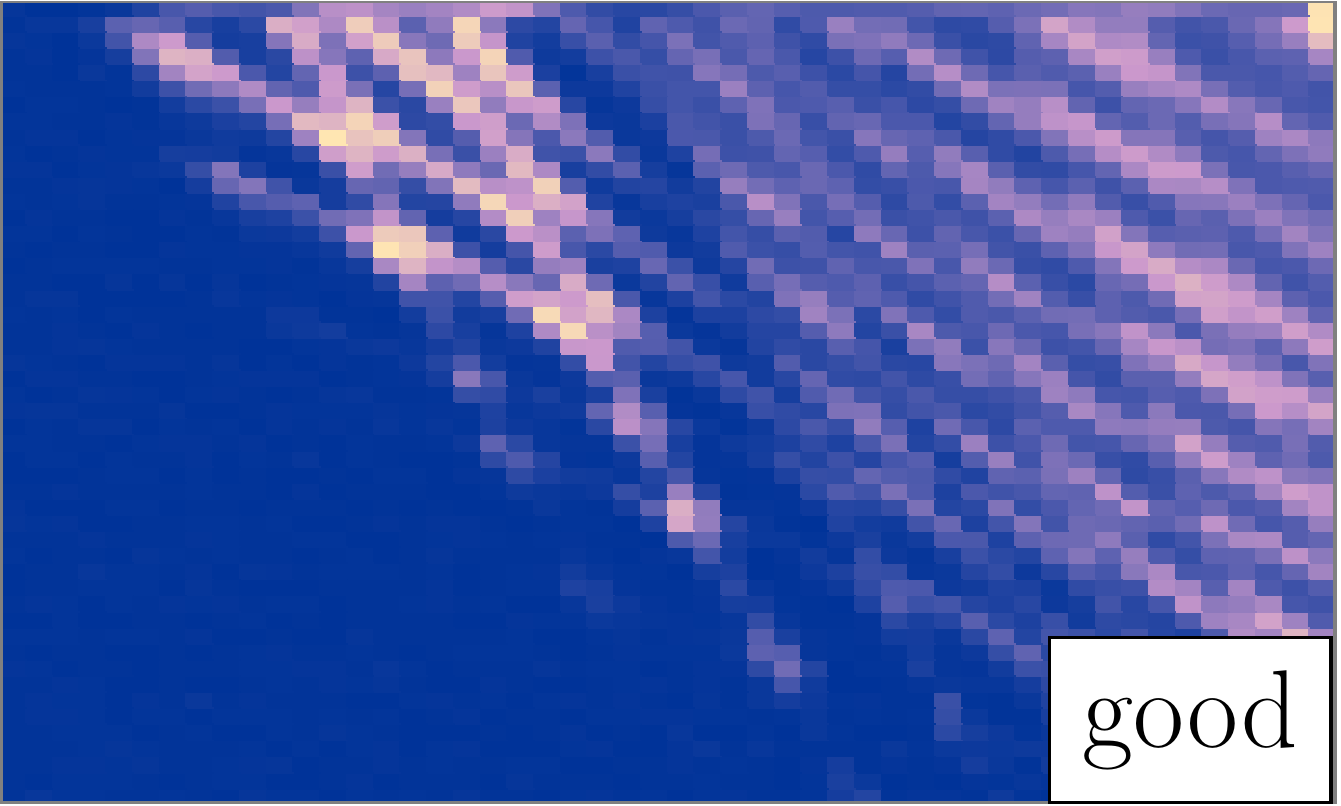}

        } 
        \sidesubfloat[]{%
        \label{fig:good_single}
            \vspace{-2in}
            \hspace{-0.1in}%
                                                                \noindent\stackinset{l}{1pt}{b}{0pt}{ 
                      {\setlength{\fboxsep}{1pt}\colorbox{white}{\includegraphics[width=0.045\columnwidth]{cartoons-37}}}}
                        {          \includegraphics[width=0.2\columnwidth]{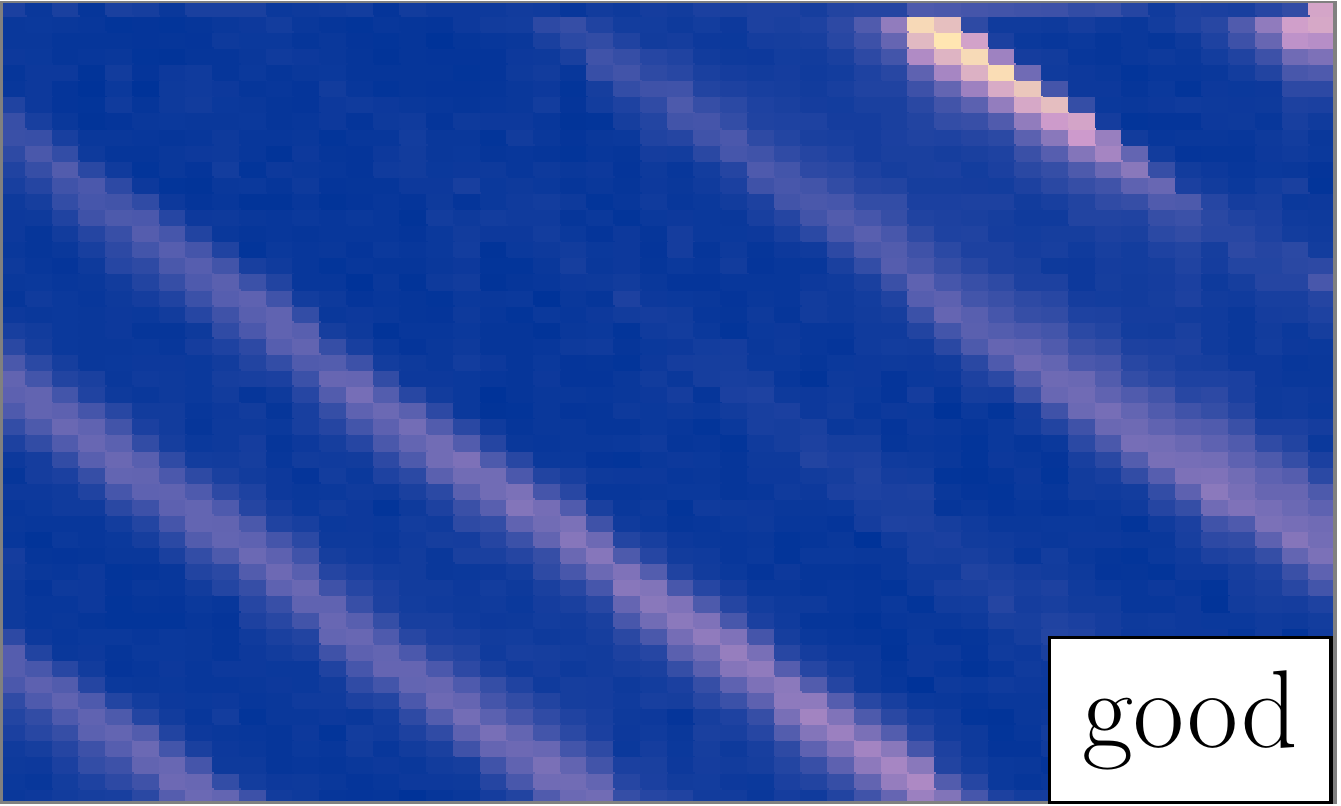} }
            \includegraphics[width=0.2\columnwidth]{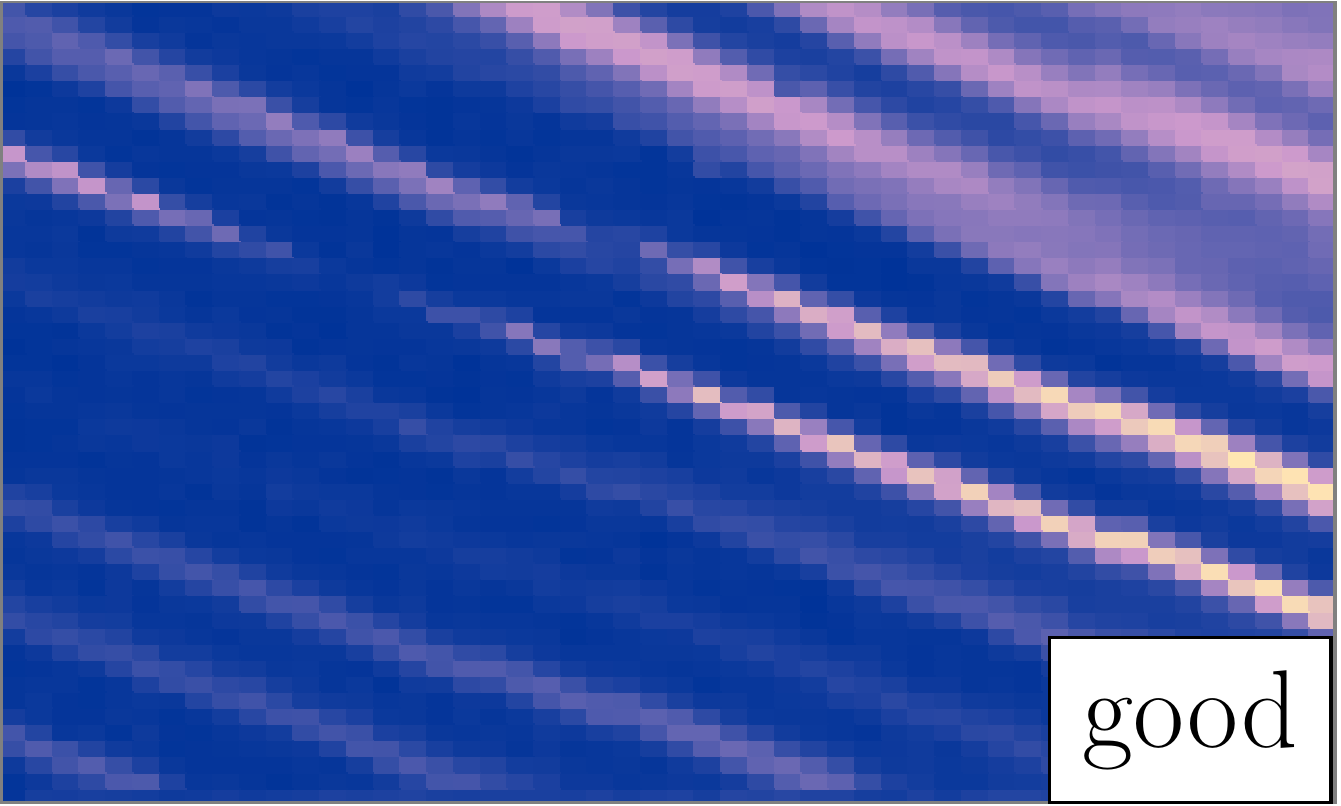}
            }
                \\
        \sidesubfloat[]{%
        \label{fig:noisy_exp_double}
        \vspace{-2in}
        \hspace{-0.1in}%
                                                            \noindent\stackinset{l}{1pt}{b}{0pt}{ 
                      {\setlength{\fboxsep}{1pt}\colorbox{white}{\includegraphics[width=0.045\columnwidth]{cartoons-37}}}}
                        { \includegraphics[width=0.2\columnwidth]{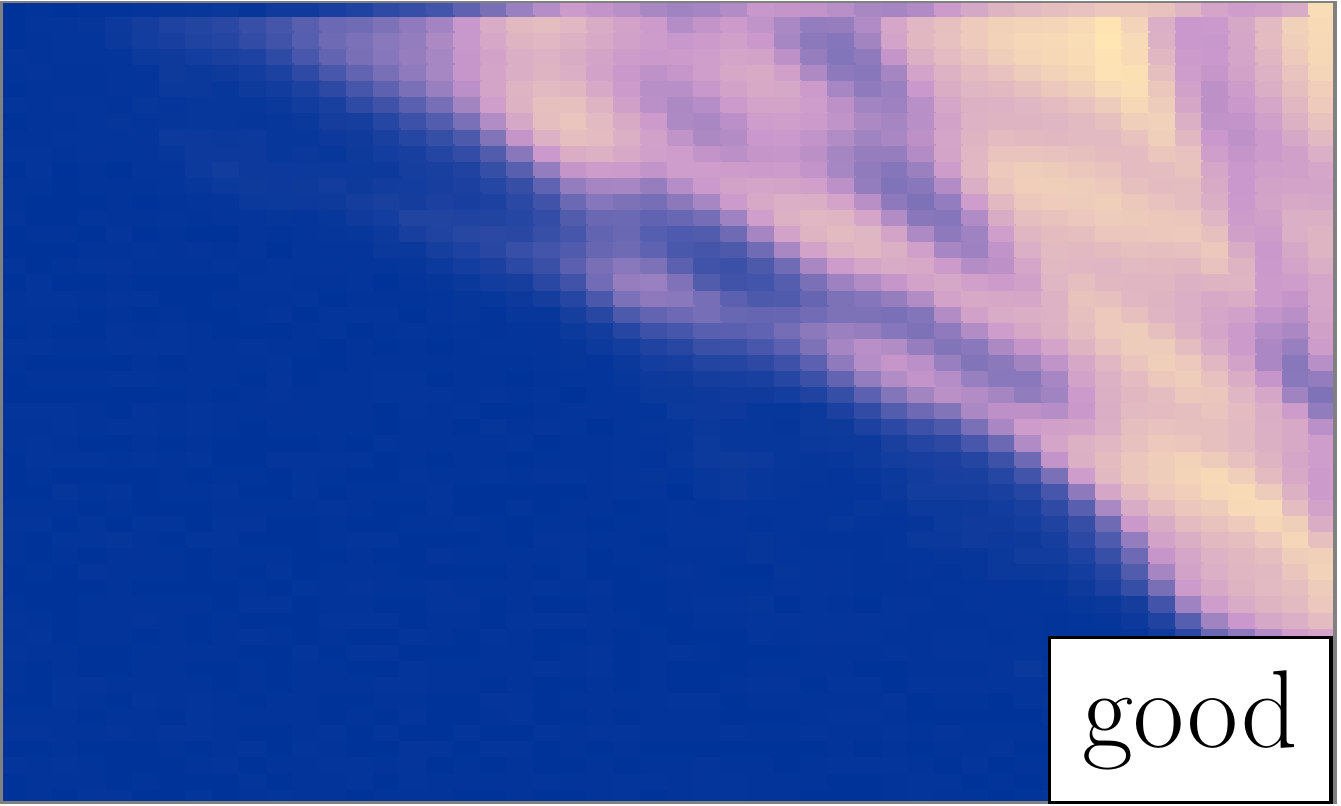}}
            \includegraphics[width=0.2\columnwidth]{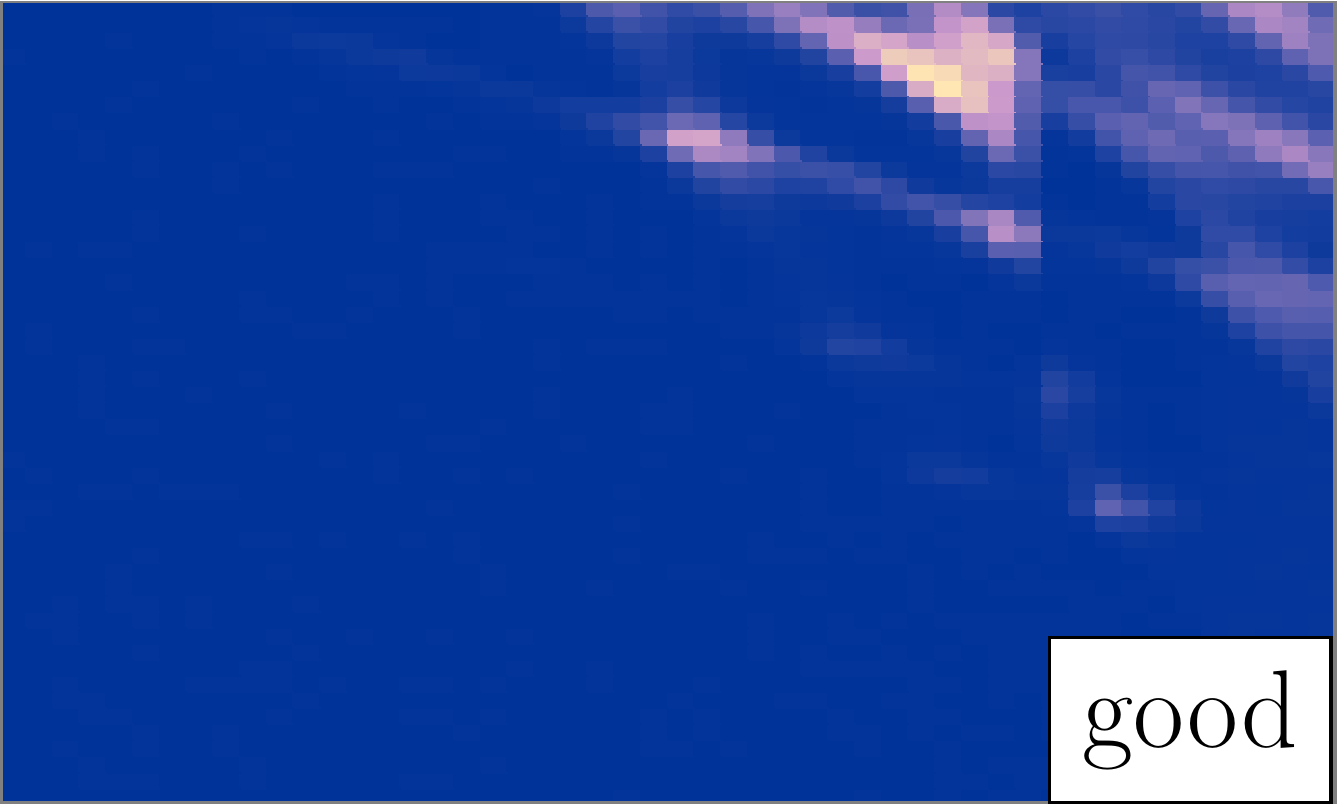}
        } 
        \sidesubfloat[]{%
        \label{fig:noisy_exp_single}
            \vspace{-2in}
            \hspace{-0.1in}%
                                                                \noindent\stackinset{l}{1pt}{b}{0pt}{ 
                      {\setlength{\fboxsep}{1pt}\colorbox{white}{\includegraphics[width=0.045\columnwidth]{cartoons-37}}}}
                        {            \includegraphics[width=0.2\columnwidth]{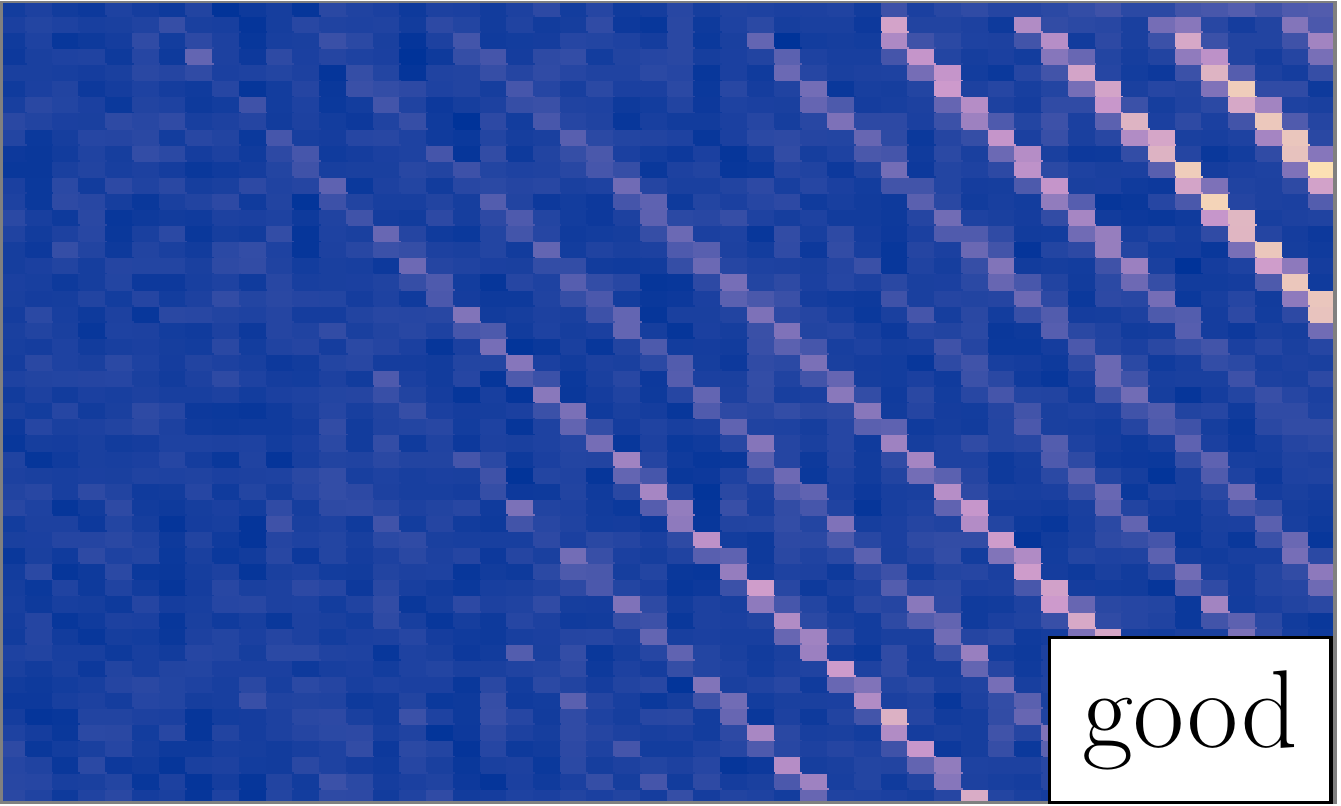} }
            \includegraphics[width=0.2\columnwidth]{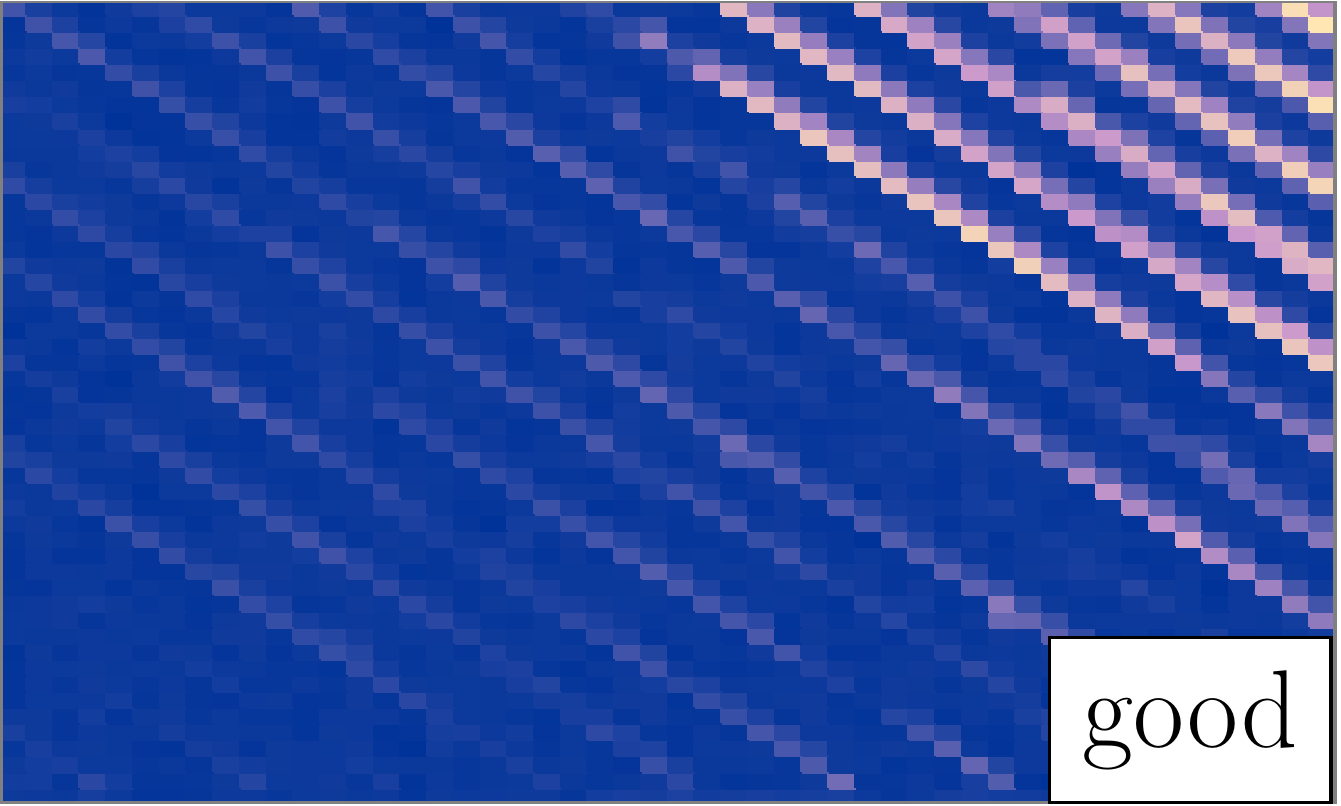}
        }
    \end{tabular}
    \caption{Examples of ideal and good experimental data, as well as noise types typically seen in measurements. Insets indicate which device type, either nanowire or two dimensional electron gas, the quantum dots were formed in. 
            (a) Ideal double dot charge stability diagram formed in a InSb nanowire. Both diagrams show similar features as our synthetic data, such as isolated triple points or a honey comb structure of weak and strong coupling regimes respectively. 
            (b) Ideal single dot charge stability diagrams formed in a GaAs two dimensional heterostructure featuring sharp and regular charge transitions. 
            (c) Good double dot charge stability diagrams showing easily identifiable triple points suitable for further fine tuning.
            (d) Good single dot charge stability diagrams showing broadened and weak transition lines.
            (e) Good double dot regimes  with pinch-off current modulation and broadened triple points (left image) and charge jumps (right image).
            (f) Good single dot regimes illustrating noise models, with white noise, $1/f$ noise and random current modulation (left image) as well as pinch-off current modulation (right image).
        }
    \label{fig:exp_data}
\end{figure*}

%


Our experimental data originates from quantum dots formed in InSb nanowires \cite{Kroll2019} as well as GaAs two dimensional electron gases \cite{Croot:2018iq, Darulova2020}. Each charge stability diagram is hand labelled by two labels indicating the tuning regime, i.e single or double, and quality, i.e.sufficient or insufficient for subsequent tuning steps.
Diagrams are labelled as sufficient if they feature clear triple points suitable for qubit parameter fine tuning procedures discussed in Ref. \cite{Teske2019, Botzem2018, VanDiepen2018a}, and insufficient otherwise.

The sufficient data is further divided into ideal and good measurements, and we assess the classification accuracy on the subsets 'ideal', 'good', and 'all' measurements. Ideal measurement outcomes show features similar those found in synthetic data. Good measurements diagrams show some types of noise, but are suitable for further fine tuning. We have 221 ideal, 1681 good and 4613 bad charge stability diagrams, examples of which are shown in \autoref{fig:exp_data}. 
The data also shows common noises such as a gradual current drop towards negative gate voltages, broadening of transitions and charge jumps, which are illustrated \autoref{fig:noisy_exp_double} as well as white noise, $1/f$ noise  and random current modulation, displayed in  \autoref{fig:noisy_exp_single}.


 \begin{figure}[!t]
 \includegraphics[width=\columnwidth]{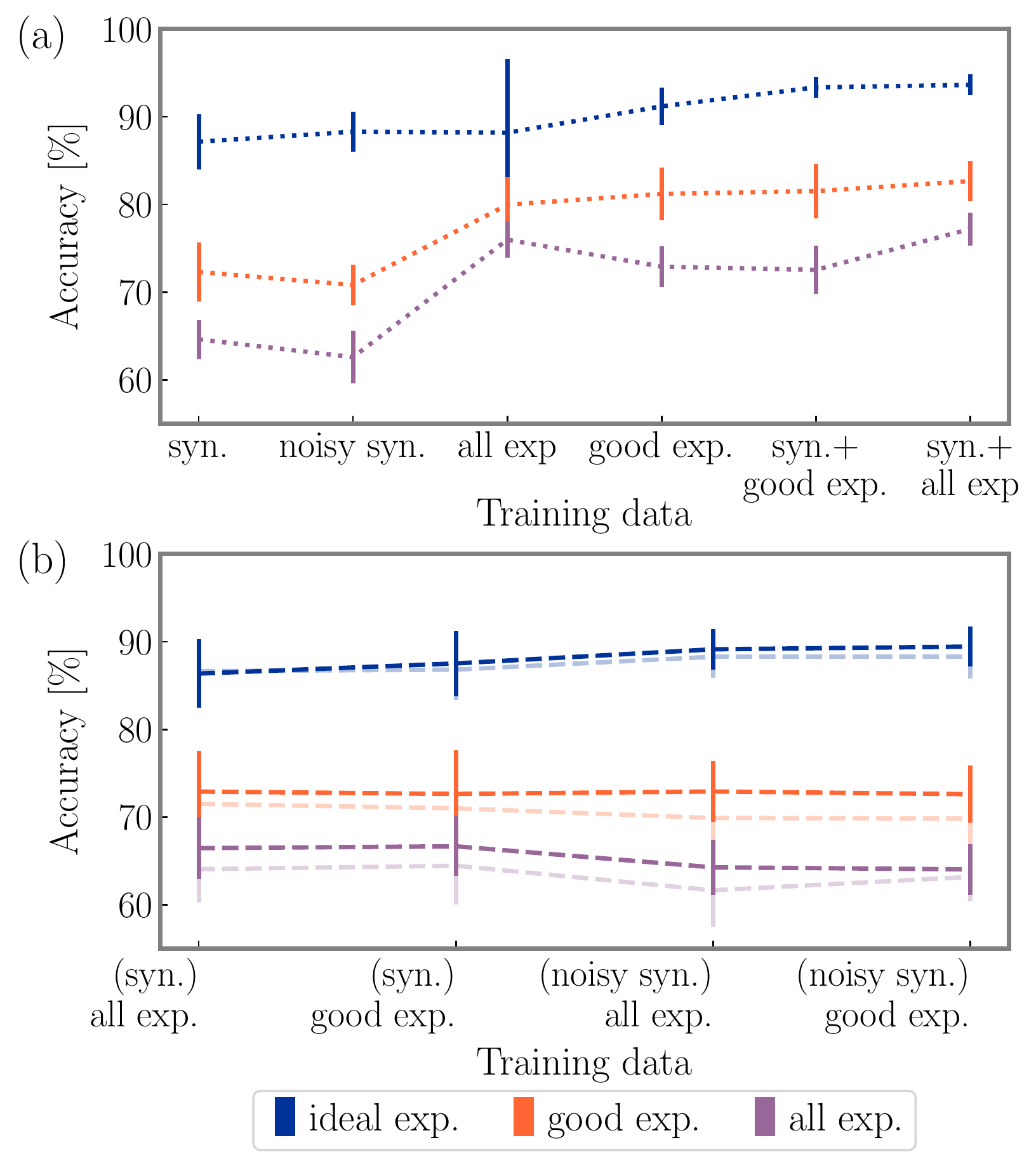}
  \caption{Neural network classification results showing average accuracy and standard deviation over ten train and test splits. Dashed lines are a guidance to the eye and connect average values, vertical bars indicate standard deviations. Classification is performed on ideal, good and all experimental (exp.) data.
            (a) Classification accuracies when trained on either synthetic (syn.), noisy synthetic (noisy syn.) or a combination of synthetic and experimental data. 
         (b) Classification accuracies achieved using transfer learning. Pre-training data is indicated in parenthesis. Pre-trained and post-transfer accuracies are plotted in light and dark coloured lines respectively. }
     \label{fig:results_plot_my_CNN}
 \end{figure}

  \begin{figure}[!t]
 \includegraphics[width=\columnwidth]{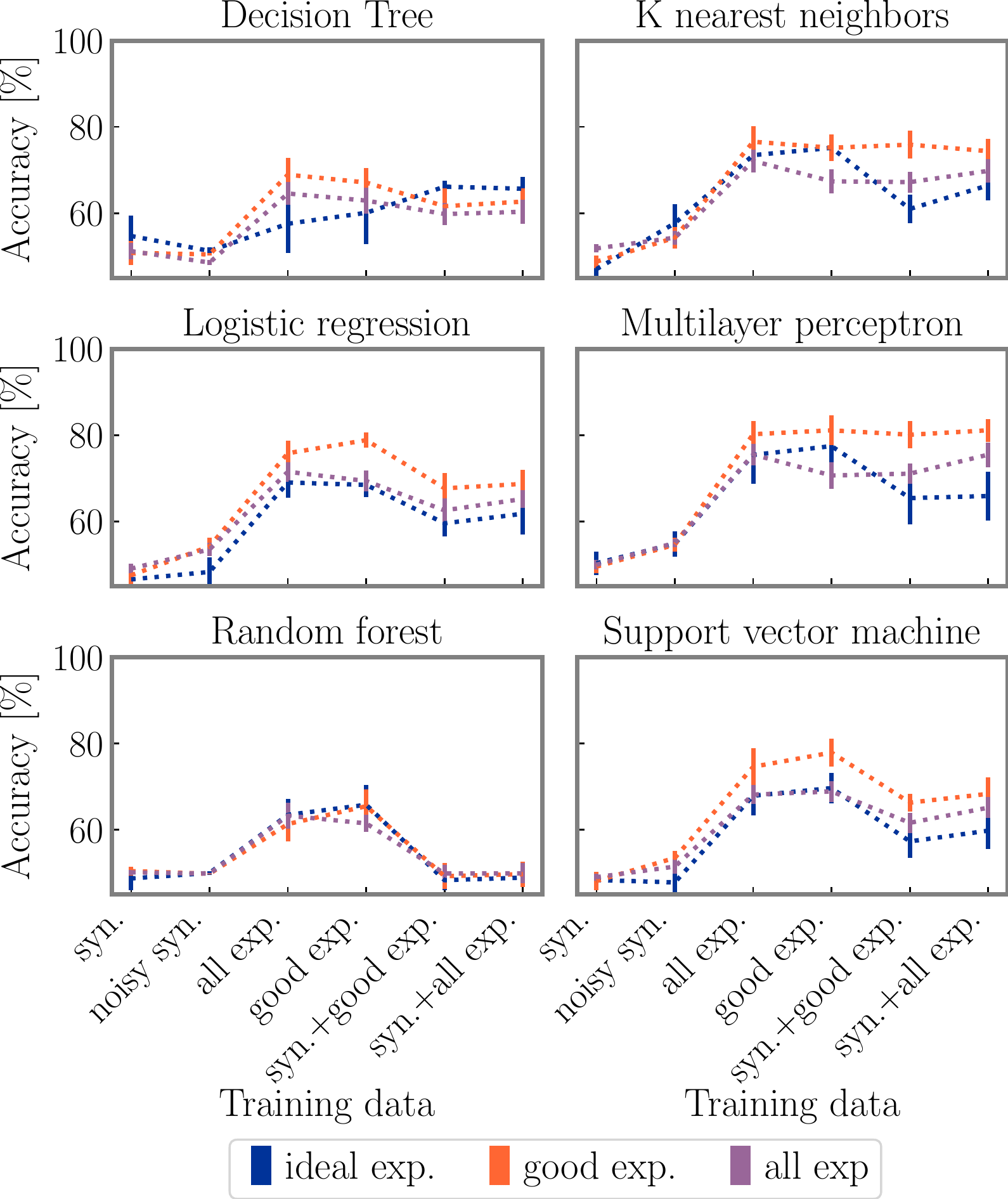}
  \caption{Binary classifier accuracies showing average accuracies and standard deviation over ten train and test splits. Dashed lines are a guidance to the eye and connect average values, vertical bars indicate standard deviations. Synthetic data or combinations of synthetic and experimental data was used for training and ideal, good or all, i.e. ideal, good and poor, experimental data for testing.}
     \label{fig:binary_results_fig}
 \end{figure}

The data described above is used to assess the ability of convolutional neural networks and parametric binary classifiers to generalise from synthetic data to a variety of experimental data.
Each classifier is trained on the following dataset combinations:  noiseless synthetic data, noisy synthetic data, good experimental data, all (ideal, good and bad) experimental data, noiseless synthetic data and good experimental data, noiseless synthetic data and all experimental data.
Classification accuracies are evaluated on ideal, good and all experimental data. When applicable, these dataset are split into 80~\% train and 20~\% test set and none of the data used in training is used in testing.
We perform ten random train and test splits and report the average accuracy and standard deviation. These datasets are balanced, meaning they contain the same number of single as double quantum dots. This results in different sizes of train and test data for each dataset combination.

The results illustrated in \autoref{fig:results_plot_my_CNN} and detailed in \autoref{tab:results}  show that training the neural network on only synthetic data allows to predict ideal experimental data with an average accuracy of  87.15~\%. Broadening the scope to include good and all experimental data sees the accuracies decrease to 72.30~\% and 64.62~\% respectively.
These accuracies improve to  91.18~\% when good experimental data is used for training, and are highest when synthetic and experimental data is combined to a single dataset, reaching 93.63~\%.
 
Overall, adding synthetic data to an experimental training set improves accuracy by up to 3~\%. 
Confusion matrices for each classification, detailing which subclasses tend to be misclassified, \autoref{tab:my_CNN_conf_mat}, show that single dots tend to be misclassified as double dots more often than double dots as single dots.

Adding noise to the synthetic training data generally results in lower accuracies than noiseless synthetic data. A detailed study of the effect of individual noise models can be found in \autoref{tab:noise_results}, where only one noise type was added to the training data at a time with various amplitudes.  Only random telegraph, $1/f$  and white noise added is able enhance ideal experimental data classification by 1-2~\% when added with specific amplitudes.  
We also note high training accuracies when synthetic data is present in the training set (see \autoref{tab:results}), indicating that noise and random fluctuations are learned as concepts, called overfitting.
Training a shallower neural network with fewer weights and which is thus less prone to overfitting suffers from this as well, as can be seen in \autoref{tab:results_shallow_cnn}. Adding noise or experimental data to synthetic training data reduces training accuracy.


We further investigate benefits of transfer learning, during which the network is pre-trained on one dataset and then re-trained using a second dataset while keeping weights of all but the last layer fixed. This technique has potential to reduce training time or increase accuracies when not enough training data is available \cite{Weiss2016}.
Results of transfer learning using either synthetic or noisy synthetic data for pre-training and good or all experimental data for re-training are illustrated in \autoref{fig:results_plot_my_CNN} and detailed in  \autoref{tab:transfer_results}.
We see little improvement when predicting ideal experimental data, but networks predicting good and all experimental data benefit from transfer learning by up to 5 \% compared to the pre-trained network. 
Overall, these accuracies are up to 5 \% lower than when training with both datasets together once.


Classification accuracies of the binary classifiers studied are summarised in \autoref{fig:binary_results_fig} and detailed in \autoref{tab:binary_results}. All binary classifiers show lower accuracies than the neural network, ranging between 50~\% and 80~\%.
The classification accuracies is highest when trained with experimental data only. The exceptions are the decision tree classifier predicting ideal experimental data, and k-nearest neighbour and multi-layer perceptron predicting all experimental data, which benefit from added synthetic training data.
Unlike the neural network, training with ideal experimental data does not show higher accuracies than good or all experimental data combined. The multi-layer perceptron performs best, followed by K-nearest neighbour, logistic regression and support vector machines. Adding noise to synthetic training data increased accuracies for the multi-layer perceptron, k-nearest neighbour and logistic regression, while it decreases accuracies for the decision tree classifier and support vector machine. 
Similar to the neural network, these classifiers have high training accuracies, suggesting that overfitting occurs when trained only on synthetic data.


To summarise, we find the highest prediction accuracies are achieved by training these classifiers on either experimental or a combination of synthetic and experimental training data.
Adding only a small experimental dataset to a large synthetic dataset allows the convolutional neural network to learn the specific type of noise present in real measurements and hence improve classification accuracy. As all models suffer from overfitting, adding more variety to training data from either improved device models or more experimental data is necessary to achieve higher success rates. 
Particularly, neural networks with additional convolutional layers could reach higher accuracy and learn a larger variety of charge stability diagrams originating from different materials and device architectures. But a deeper architecture is more prone to overfitting and requires an even larger training dataset.
Segmenting experimental data into regions with fewer regime variations may also increase accuracies.

The noise added to synthetic data does not significantly improve classification accuracies, showing that it does not  match the experimental data. More realistic noise models and quantum dot simulations taking into account impurities and fabrication defects are expected to improve accuracy of classifiers trained on synthetic data.

Realistic semiconductor quantum dot simulations are complex and noise encountered in today's state of the art devices, which have been used in this work, are not well understood. Future devices with less fabrication variances and impurities may reduce noise and facilitate charge state detection based on supervised machine learning using synthetic data. But until these devices are reliable and simulations sophisticated enough to reproduce their behaviour, investing time into labelling experimental data is required.
 
We thank Rachpon Kalra and John M. Hornibrook for helpful discussions and critical feedback.

The data that support the findings of this study are available from the corresponding author upon reasonable request.

%
\nocite{*}
\bibliography{literature_data_analysis}
%
\cleardoublepage

\setcounter{figure}{0} \renewcommand{\thefigure}{A.\arabic{figure}} 
\setcounter{table}{0} \renewcommand{\thetable}{A.\arabic{table}} 

\appendix

 \section{Noiseless synthetic data}\label{ax:syn_data}
 
Our synthetic dataset of simulated single and double dot charge stability diagrams is based on data generated by a capacitance model \cite{VanderWiel2003} and the Qflow-lite dataset \cite{zwolak_qlite}, which uses the Thomas-Fermi approximation. Examples of both dot regimes generated by these models are shown in \autoref{fig:syn_examples}.
The capacitance model replicates a device made of six gates coupled to two dots, similar to device architectures used  to define charge and spin qubits \cite{Yoneda2020, Croot:2018iq, Hendrickx2020, Cerfontaine2020, Watson2018, Xue2019, Harvey2018}. 
A set of 2000 diagrams was generated by randomly sampling capacitances from a Gaussian distribution centred around one of several capacitance combinations generating diagrams encountered in experiments. 

We use segments of the original Qflow-lite dataset made available online, divided into 15 subregions of 30 x 30 pixels per original charge stability diagram. We use python's scikit-image\cite{scikit-image} resize method to resize these segments to 50 x 50 pixels, the size of our data.
We first normalise each diagram individually to a range between 0 and 1 and then multiply them by an overall factor of 0.3 to ensure charge transitions to be of similar  strength as in our experimental and synthetic data.
The original Qflow-lite labels, which are vectors of four components indicating the probability of the diagram showing a fully open, fully closed, single dot and double dot regime, were transformed into binary labels differentiating only between single and double dot. For this, only diagrams with a fidelity of being in either single or double dot regime of more than 60 \% are retained and relabelled as either a single or double dot.

 \section{Noisy synthetic data}\label{ax:syn_noise}
 
 We implement five noise models typically encountered in experiments that are added to the noiseless synthetic data, which are referred to as noisy synthetic data sets. These noise types are white noise, random telegraph noise, $1/f$ noise, charge fluctuations on gates, low frequency current modulations and pinch-off current modulation. White noise typically arises due to thermal fluctuations, while $1/f$ noise and charge fluctuations on gates are two types of random fluctuations due to defects in the semiconductor. Random telegraph noise on the other hand is a low-frequency modulation of current caused by the spontaneous capture and emission of charge carriers. Low frequency current modulations and pinch-off current modulation are consequences of the electron gas being depleted for decreasing gate voltages.
Additional examples of each noise type are shown in \autoref{ax:noise_examples}. Noise is generated as follows:

The $1/f$ noise is generated in frequency domain by creating 2D frequencies mesh and taking the inverse of their norm to calculate the  magnitude of spectral coefficients:
\be
C_{k, l} = 
\begin{cases}
     \frac{1}{\sqrt{f_{k}^{2} + f_{l}^{2}}} , & \text{if $f_{k}^{2} + f_{l}^{2}>0$}.\\
    0, & \text{otherwise}.
  \end{cases}
\ee
We set the phases of these coefficients to random values:
\begin{align}
 C_{ k, l} = &  C_{ k, l} e^{i \phi_{k, l} },
\end{align}
where $\phi_{k, l}$ is chosen randomly from a uniform distribution over  $[0, 2\pi)$.
The inverse Fourier transform is then added to the images. 
White noise is generated as a 2D map of random, normally distributed coefficients with zero mean and a variance of 1. Pinch-off current modulation is achieved by convoluting the image with 
\be
W_{i, j} = \tanh(\alpha*x_{i,j}+\beta),
\ee
where  $\alpha$ and $\beta$ are drawn from a uniform distribution  between over $[0, 10)$ and $[-5, 5)$ respectively, and $x_{i,j}$ is a pixel coordinate matrix.
Random current modulation is realized by convolving an image with a 2D map of Gaussian blobs of random mean and standard deviation, drawn uniformly between $[-1, 1)$ and $[0.3, 0.8)$ respectively.
Random telegraph noise is simulated by adding charge jumps following a Poisson distribution with an expected number of occurrences drawn from a uniform distribution between $[0, 0.2)$.
Charge jumps, which appear as voltage jumps on gates, are achieved by randomly choosing a location in gate voltage space and a step size, defining the subregion of the current map which will be removed. The new image is then resized using python's scikit-image resize method to its original size.  A 2D Gaussian convolution is applied to all image to simulate thermal broadening.

We generate 10000 random noise maps for each noise type, 
 which are normalised to a range between 0 and 1.
These are then added to  noiseless synthetic data by choosing a random sub-selection of maps and random amplitudes distributed uniformly between 0 and 0.2, an amplitude range over which we saw the highest accuracy variation.
Random telegraph, white and $1/f$ noise are added to noiseless current maps, while random current and pinch-off current modulation maps are convolved. The resulting diagrams are scaled by the ratio of old and new maximum current values to ensure previous normalisations are preserved. 
Charge jumps are added to a number of charge stability diagrams determined by the random amplitude times the total number of diagrams in the target dataset. 

 \section{Classification}\label{ax:classification}
 \begin{table}[!t]
    \begin{tabular}{llc}
    \hline \hline
        layer & \multicolumn{2}{c}{details} \\ \hline 
        Conv2D &  filter & 32 \\
                & kernel size & (3, 3)\\
                & activation & relu \\ 
        Conv2D &  filter & 64 \\
                & kernel size & (3, 3)\\
                & activation & relu \\
         MaxPooling2D & pool size & (2, 2) \\
         			& strides & 2 \\
         Dense & units & 1024 \\
         	   & activation & relu \\
         Dropout & rate & 0.5 \\
         Dense & units & 512 \\
         	   & activation & relu \\
         Dropout & rate & 0.5 \\
         Dense & units & 128 \\
         	   & activation & relu \\
         Dropout & rate & 0.5 \\
         Dense & units & 2 \\
         	   & activation & softmax \\
	   \hline
		 & optimizer & Adam \\
		 & learning rate & 0.001 \\ 
         \hline \hline
    \end{tabular}
\caption{The neural network architecture used, with layer names in the left column and their respective parameters on the right. It was implemented in TensorFlow 2, with two convolutional layers (Conv2D) and four dense layers with a dropout rate of 0.5. The dense layers consists of 1024, 512, 128 and 2 neutrons (units) respectively. The RELU activation function is used in all but the last layer, which uses the softmax activation function. We use the Adam optimizer and a learning rate of  0.001.}
    \label{tab:cnn}
\end{table}

 We compare accuracies of a convolutional neural network and six binary classifier trained on synthetic and experimental charge stability diagrams. The convolutional neural network was implemented in python's TensorFlow 2 and is summarized in \autoref{tab:cnn}. It consists of two convolutional layers and four dense layers with a drop out rate of 0.5. The convolutional layer have 32 and 64 kernels respectively, all 3x3 in size. The dense layers consist of 1024, 512, 128 and 2 neurons respectively. The RELU activation function is used in all but the last layer, which uses the softmax activation function. We use the Adam optimiser and a learning rate of 0.001.
 
\begin{figure*}[!h]
    \begin{tabular}{c}
        \sidesubfloat[]{%
        \vspace{-2in}
            \includegraphics[width=0.2\columnwidth]{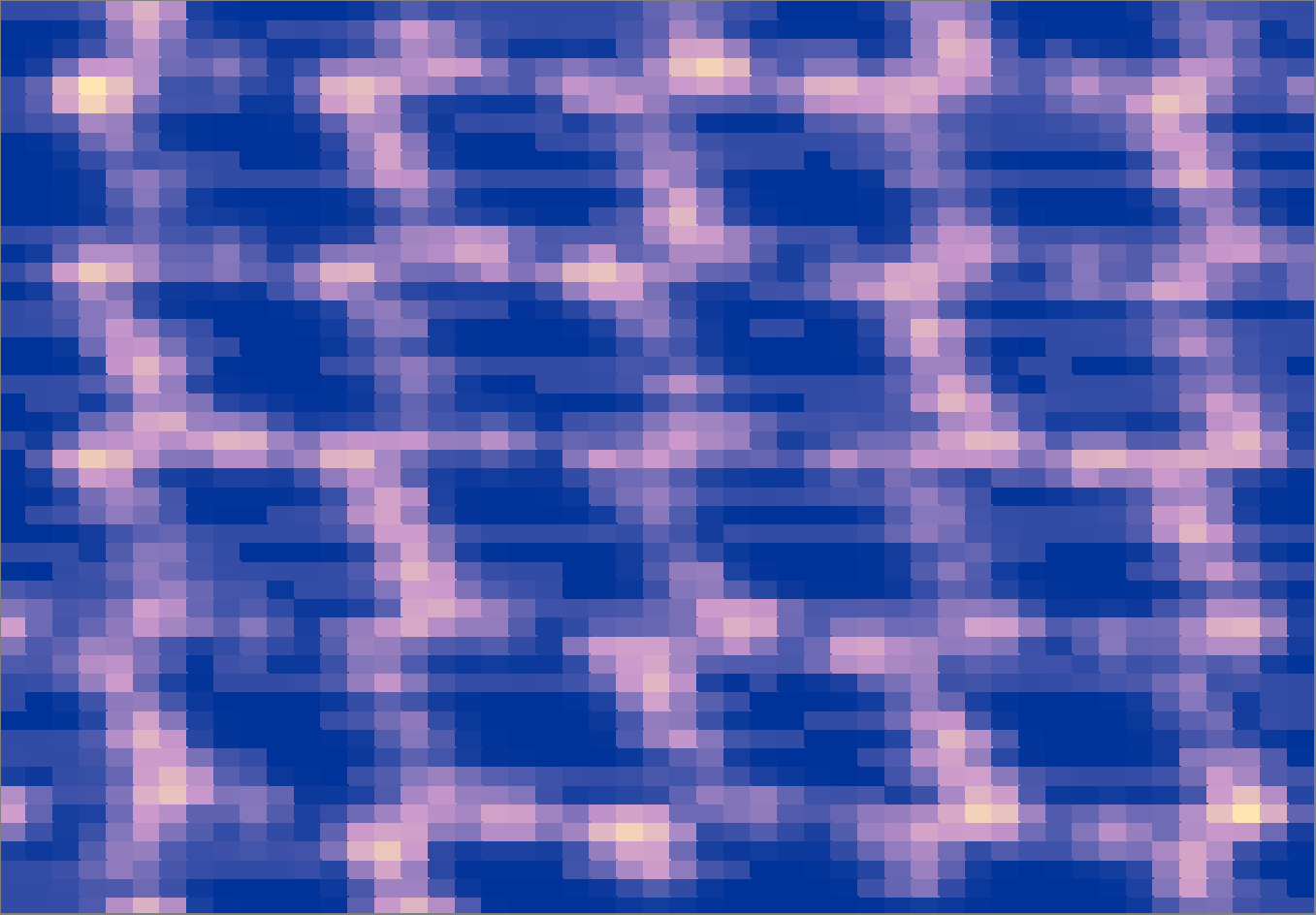}
                \includegraphics[width=0.2\columnwidth]{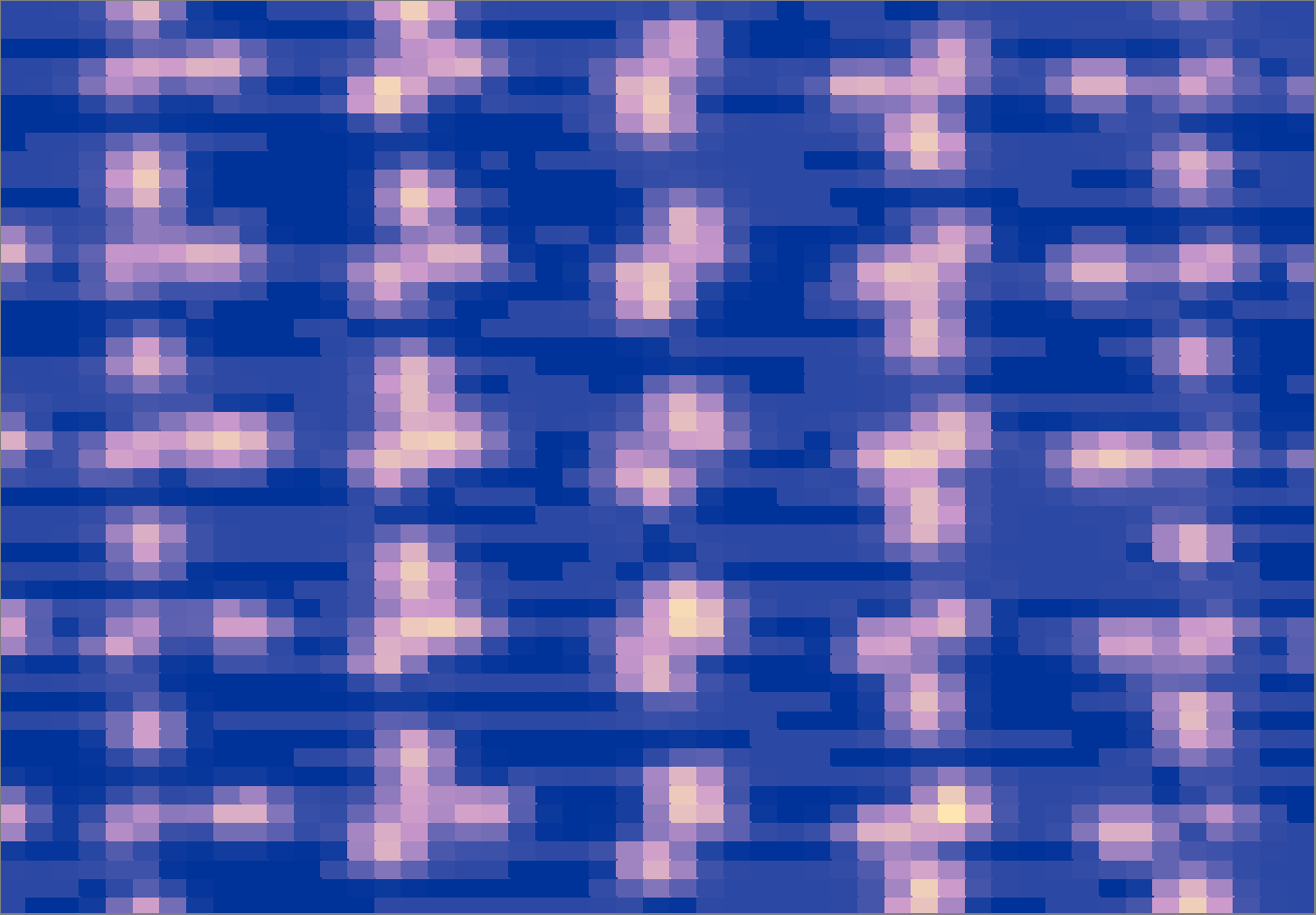}
        } 
        \sidesubfloat[]{%
        \vspace{-2in}
            \includegraphics[width=0.2\columnwidth]{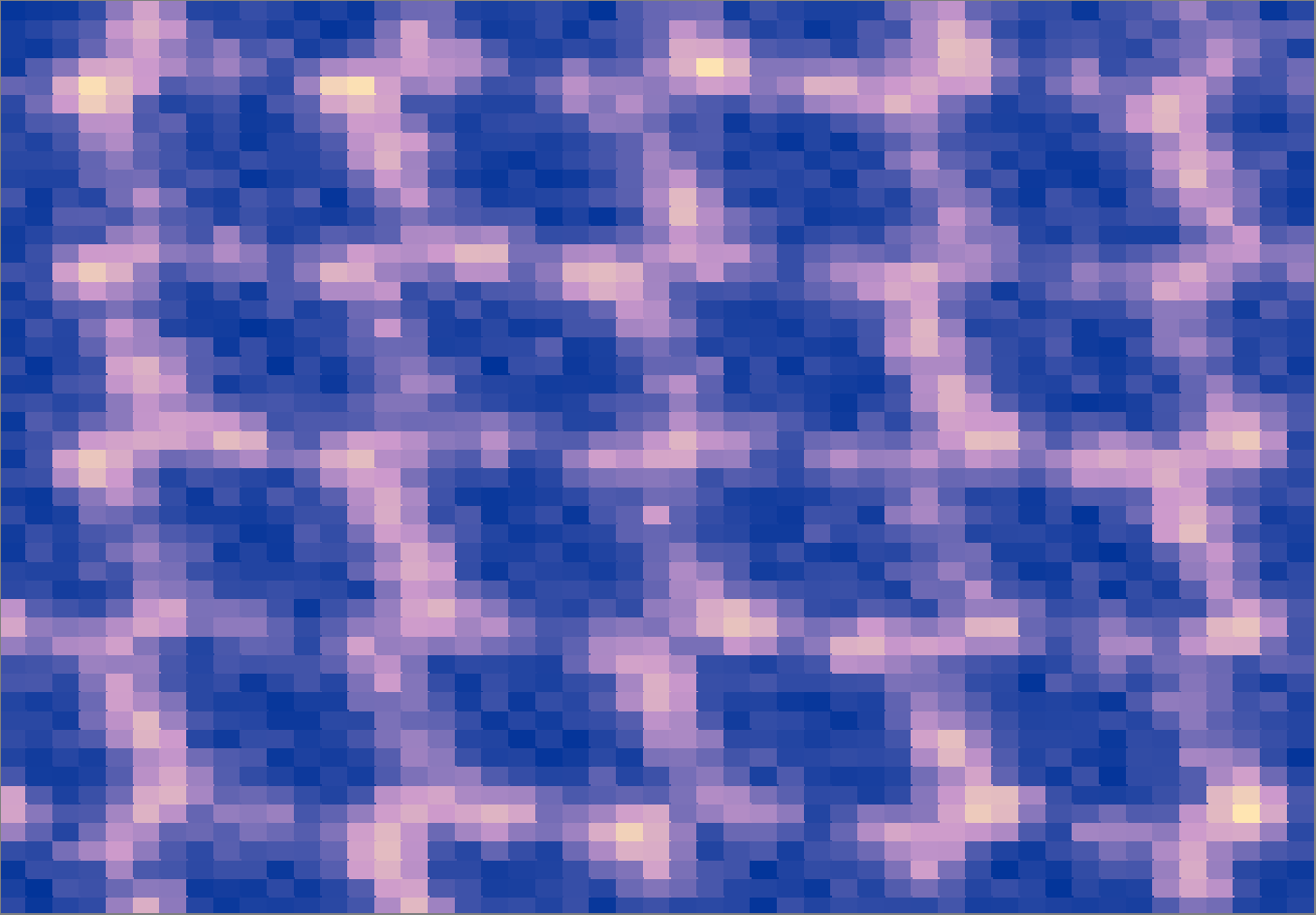}
               \includegraphics[width=0.2\columnwidth]{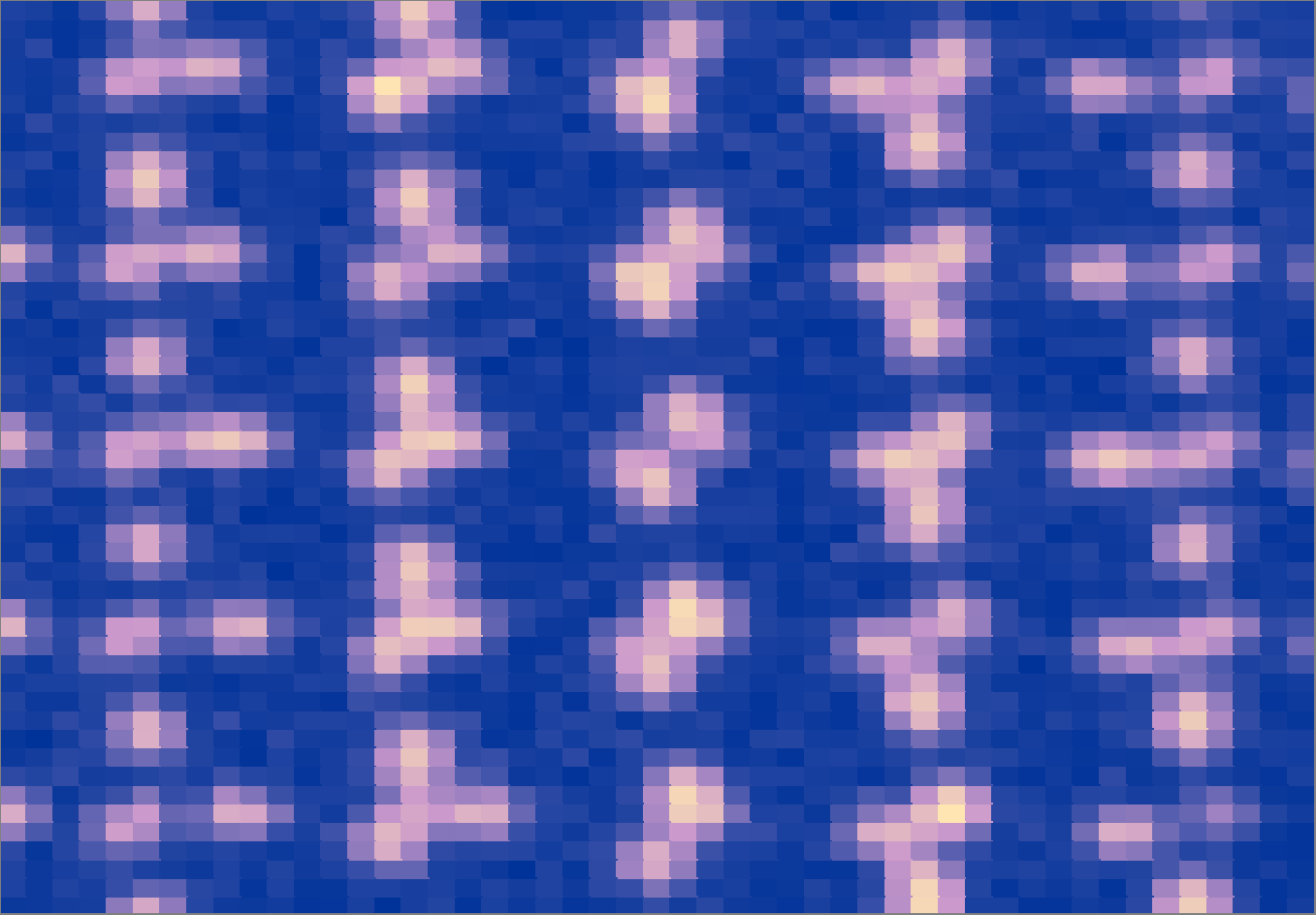}
        }
        \\ 
         \sidesubfloat[]{%
            \vspace{-2in}
            \includegraphics[width=0.2\columnwidth]{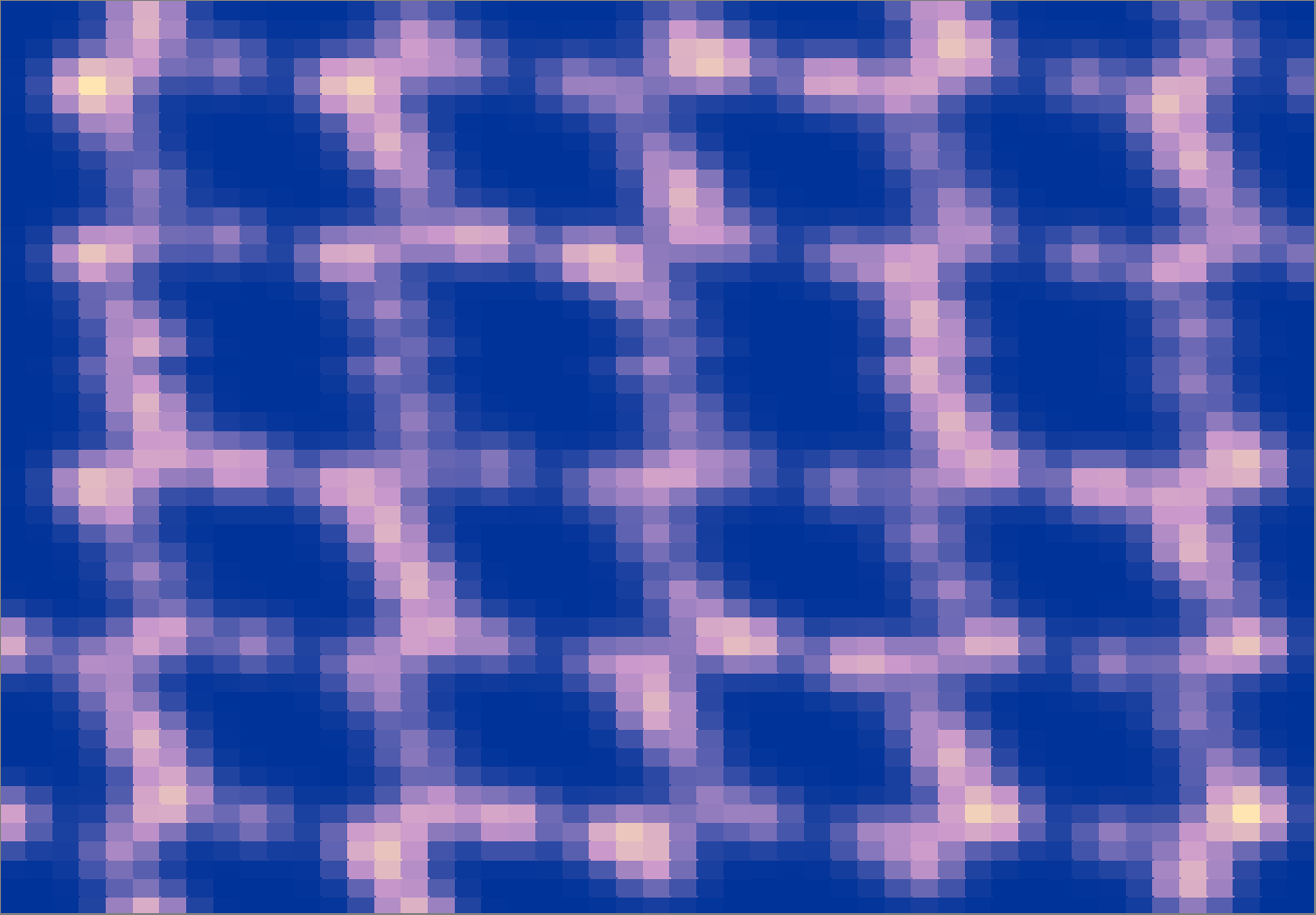}
              \includegraphics[width=0.2\columnwidth]{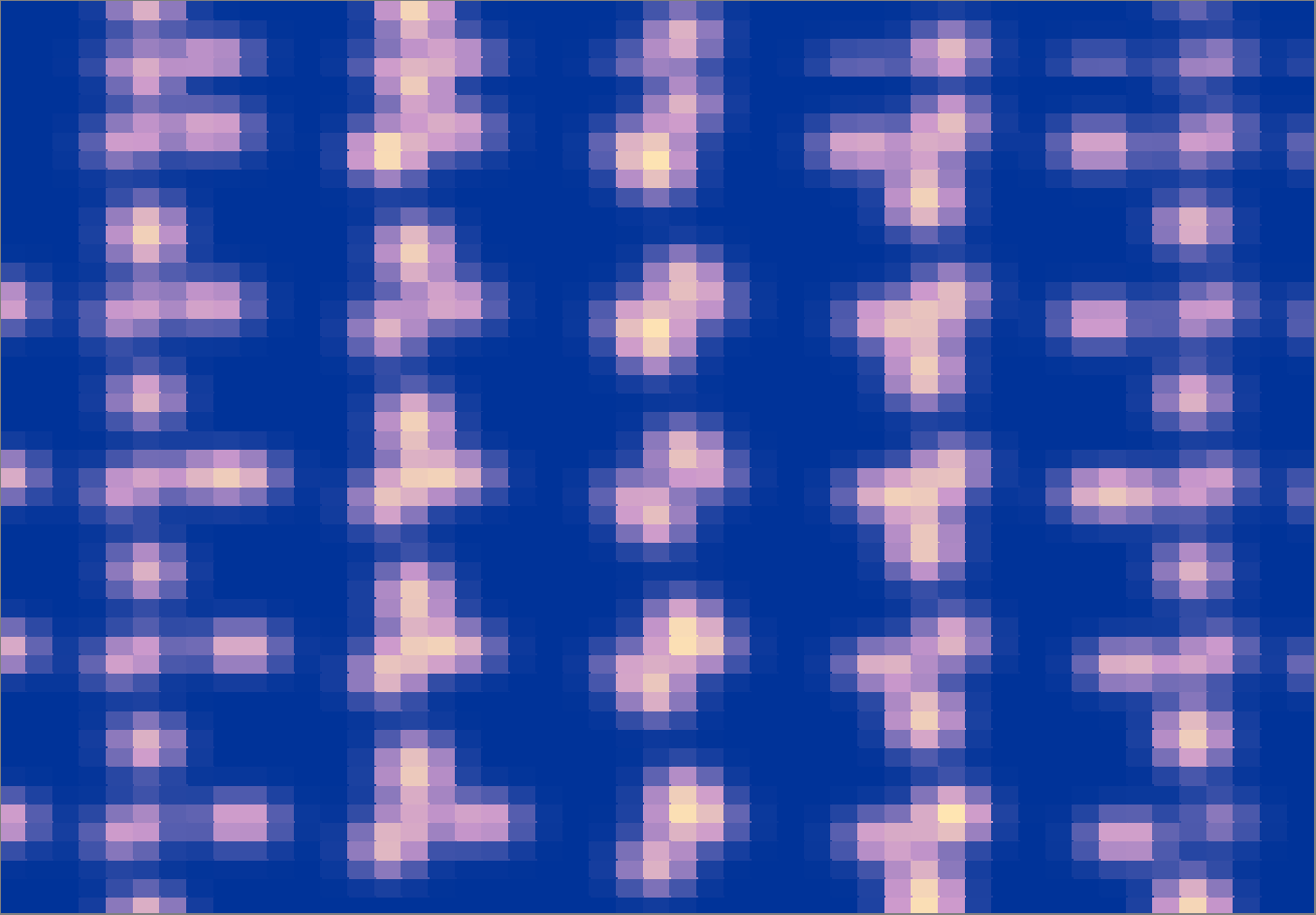}
        }
        \sidesubfloat[]{%
            \vspace{-2in}
            \includegraphics[width=0.2\columnwidth]{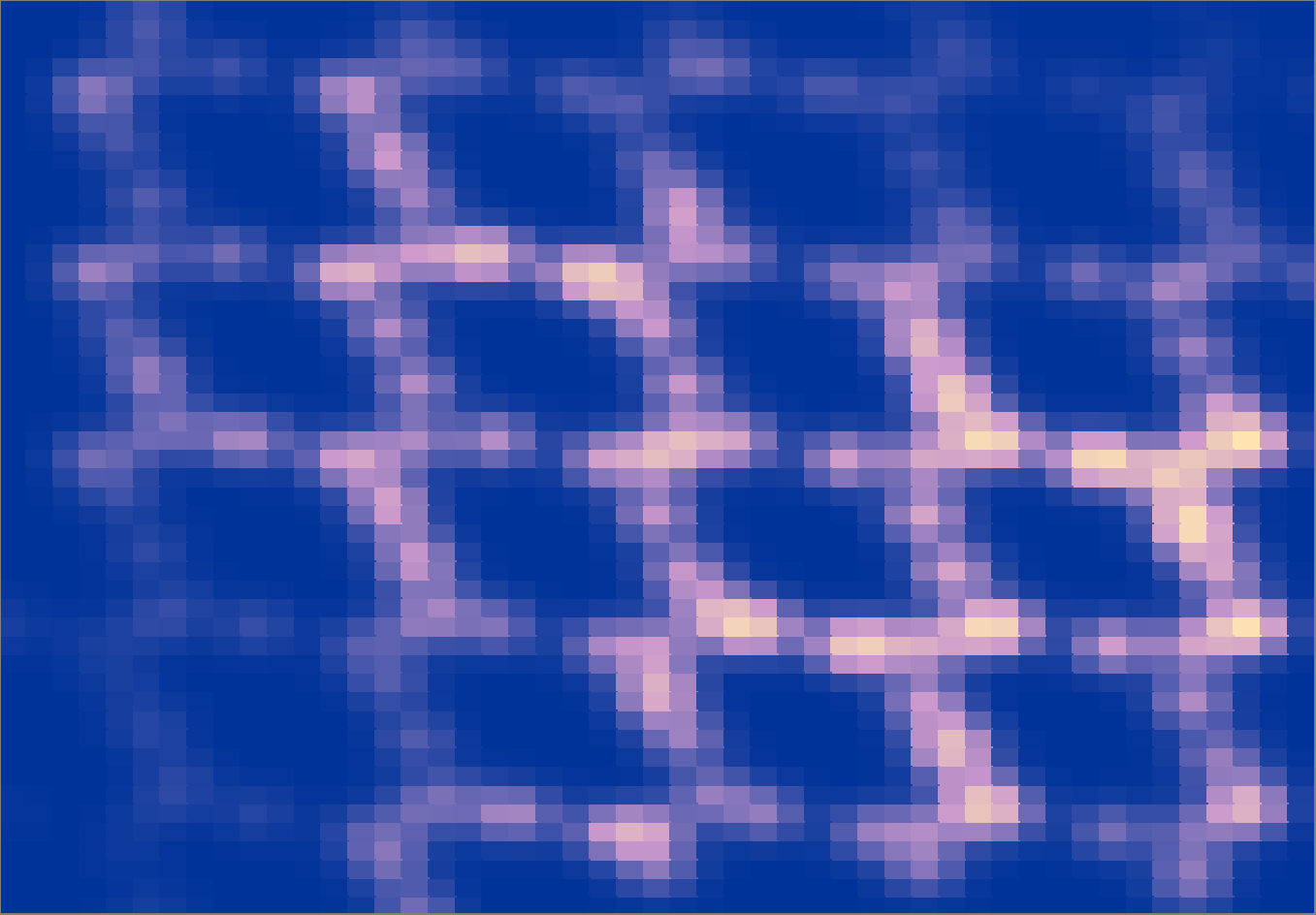} 
                        \includegraphics[width=0.2\columnwidth]{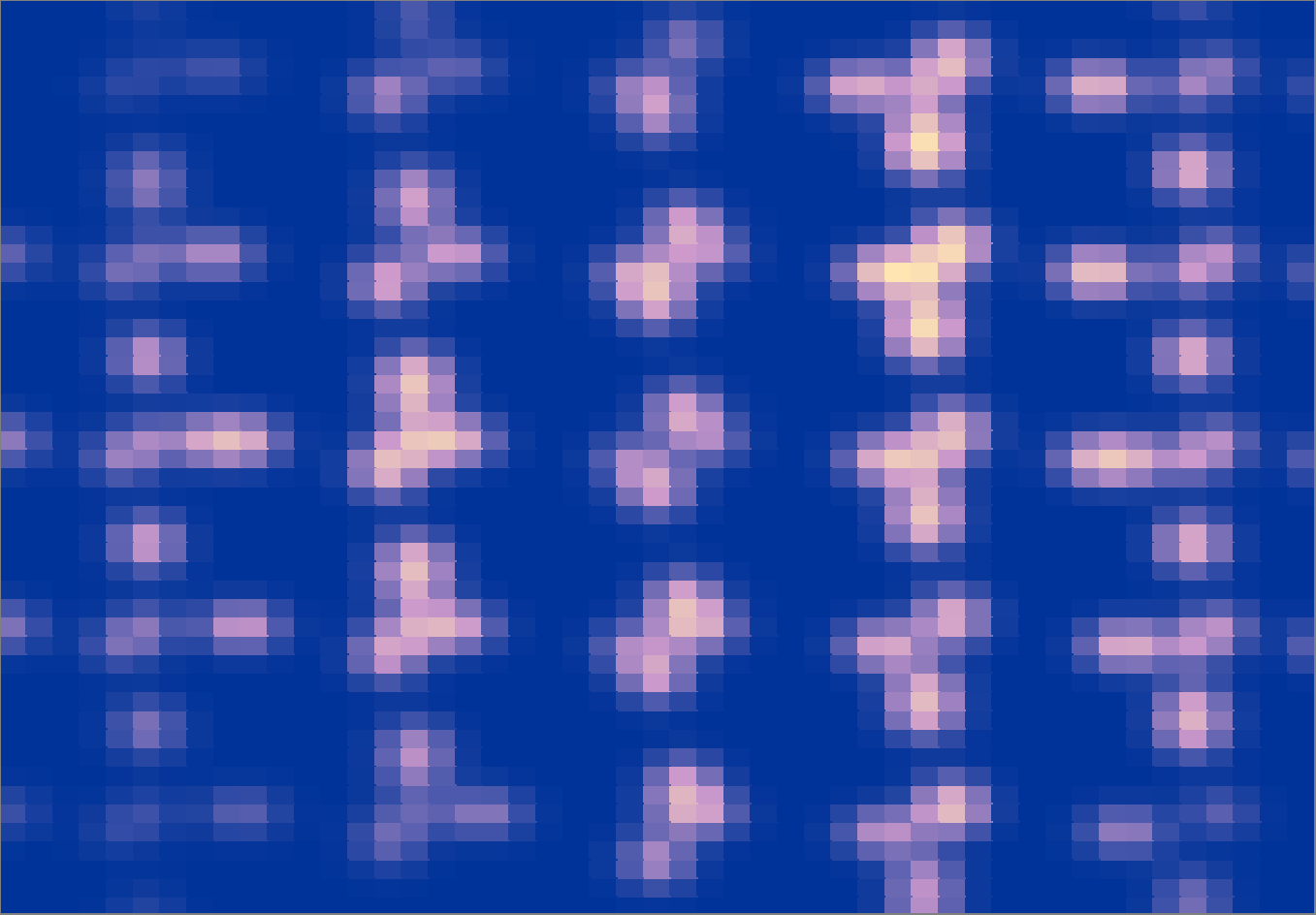} 
        }\\
        \sidesubfloat[]{%
            \vspace{-2in}
            \includegraphics[width=0.2\columnwidth]{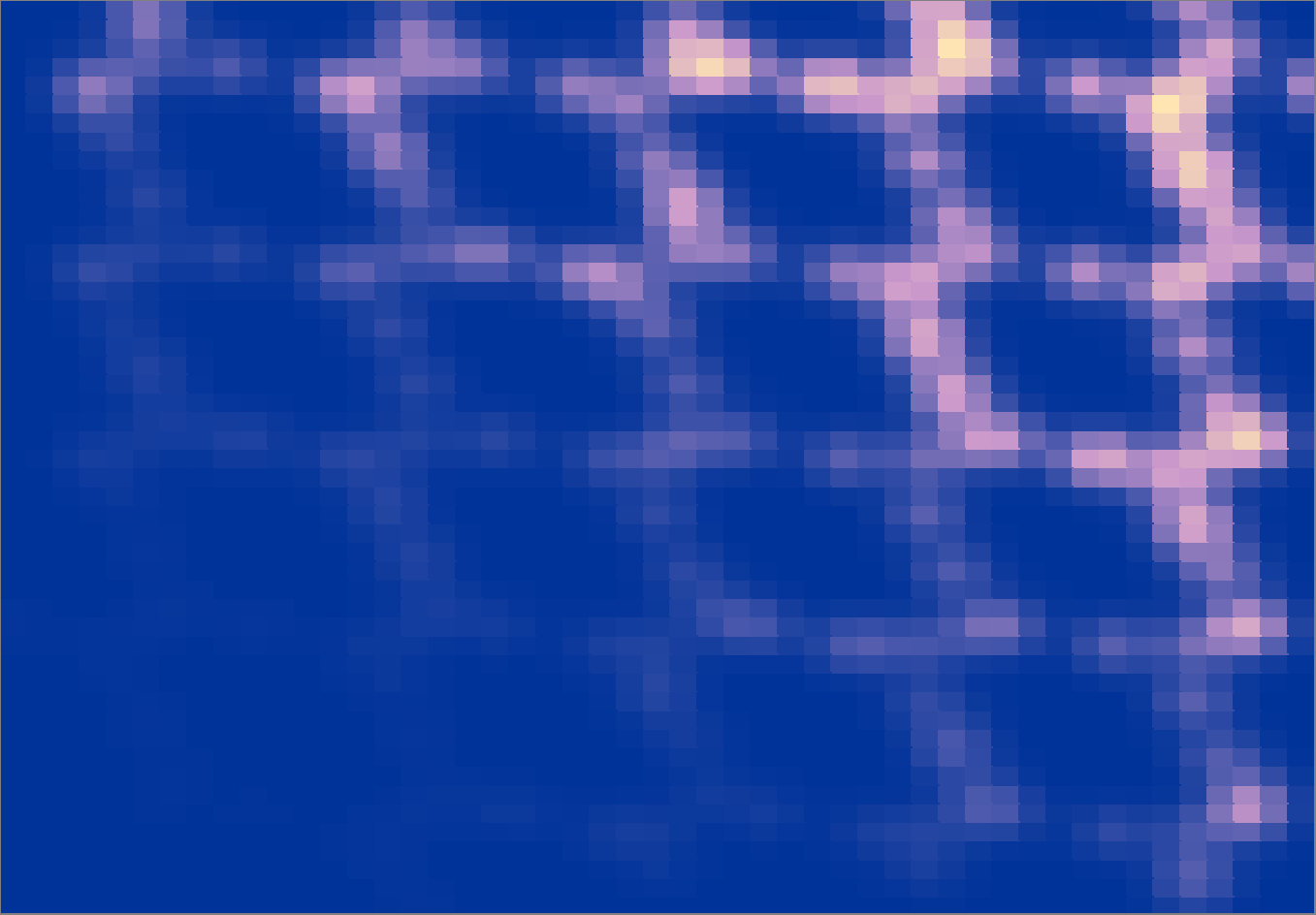}
                    \includegraphics[width=0.2\columnwidth]{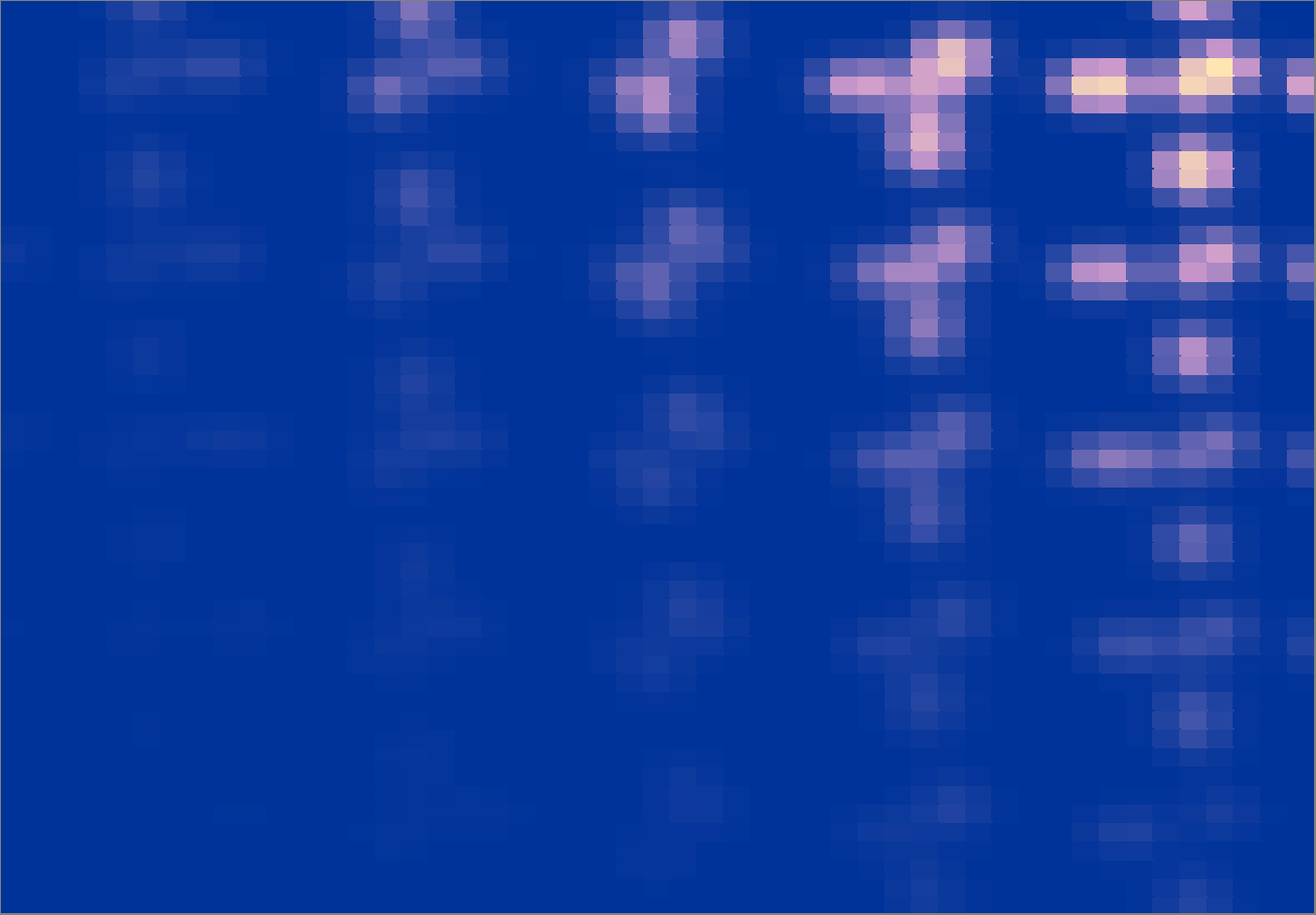}
        }
        \sidesubfloat[]{%
            \vspace{-2in}
            \includegraphics[width=0.2\columnwidth]{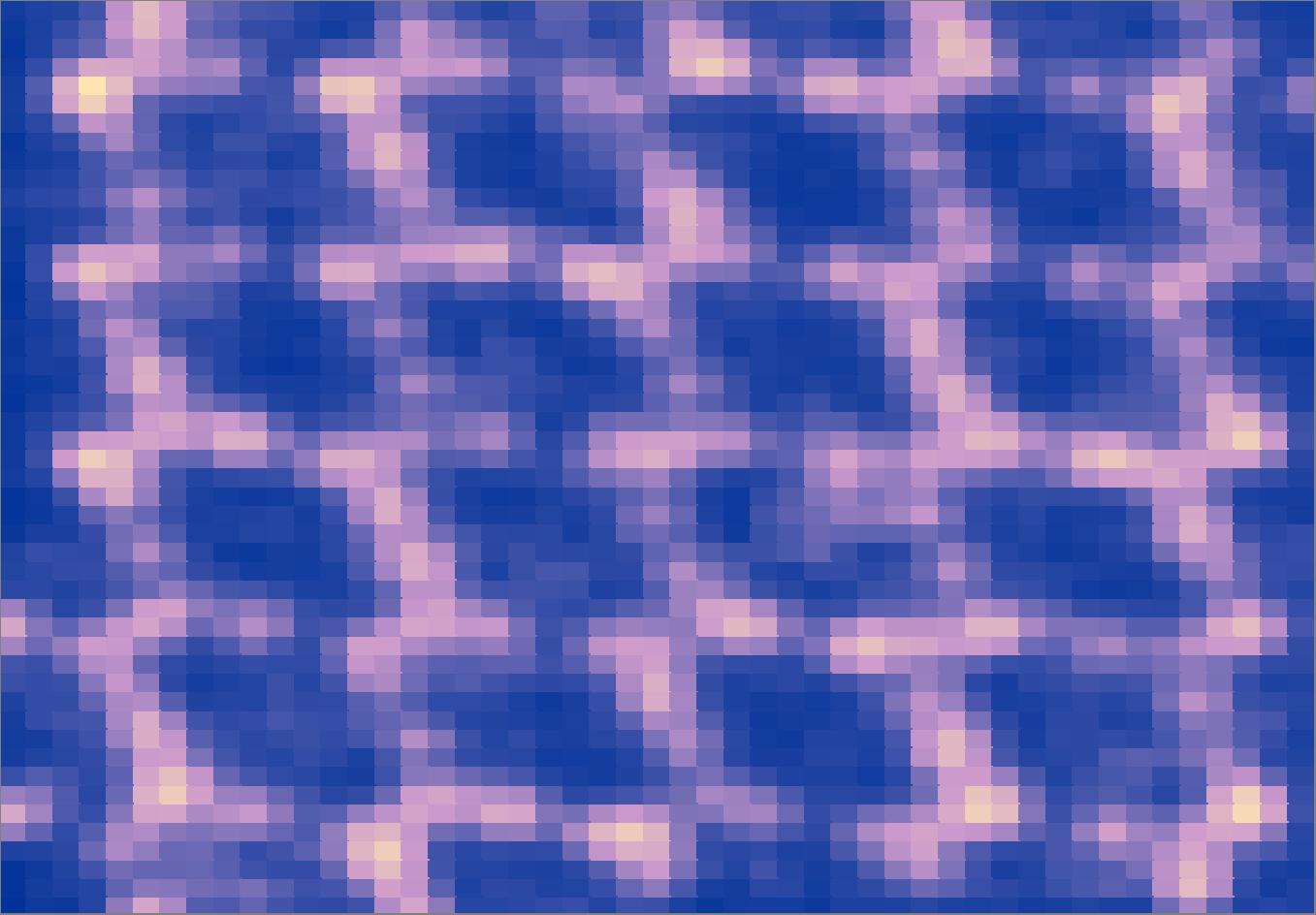} 
              \includegraphics[width=0.2\columnwidth]{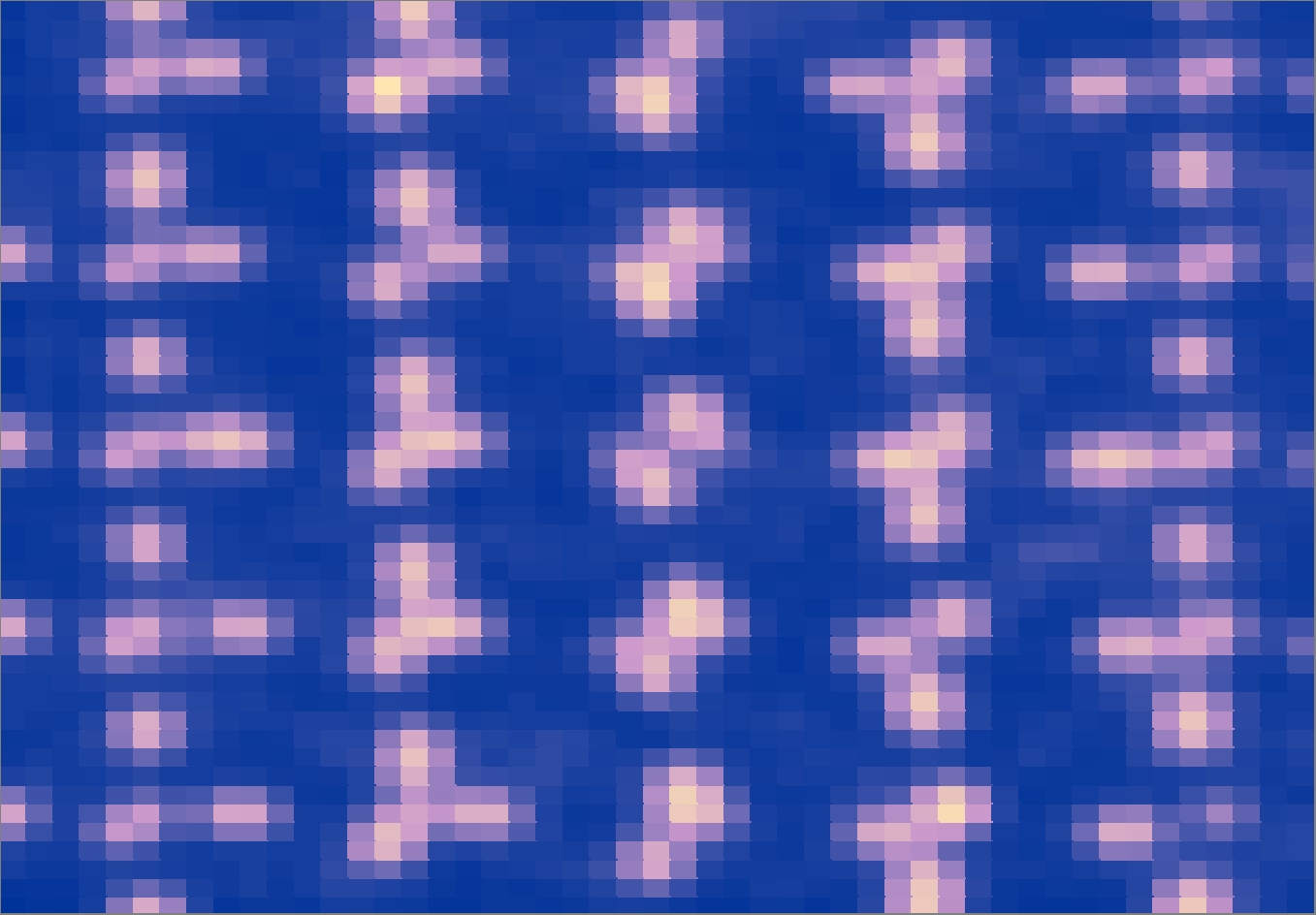} 
        }
    \end{tabular}
    \caption{More examples of noise models added to synthetic data. Random current maps were generated for each noise type independently, rescaled with varying amplitudes and then added to the noiseless synthetic dataset. 
            (a) Random telegraph noise. %
            (b) White noise. %
            (c) Charge jumps. %
            (d) Random current modulation.
            (e) Pinchoff current modulation.
            (f) $1/f$ noise.
        }
     \label{ax:noise_examples}
\end{figure*}

Additional examples of noise types added to synthetic data are pictured in \autoref{ax:noise_examples}.
 
Neural network classification results are listed in  \autoref{tab:results} and \autoref{tab:my_CNN_conf_mat}, showing accuracies and confusion matrices respectively. 
Accuracies achieved with a shallower neural network with only one convolutional layer are shown in \autoref{tab:results_shallow_cnn}.

Neural network transfer learning accuracies are summarized in \autoref{tab:transfer_results} and accuracies achieved by adding one noise type add a time to synthetic data are displayed in \autoref{tab:noise_results}.

Accuracies of simple binary classifiers are shown in \autoref{tab:binary_results}.

%

\begin{table*}[!t]
\resizebox{ \columnwidth}{!}{\begin{tabular}{@{\extracolsep{8pt}}lccccc}
\hline \hline 
  training data & \multicolumn{5}{c}{average accuracy (std)}  \\ \cline{2-6} 

{} &      training &     synthetic & clean experimental & good experimental & all experimental \\
\hline
synthetic                       &  99.22 (0.59) &  96.78 (0.43) &       87.15 (3.15) &      72.30 (3.40) &     64.62 (2.25) \\
good experimental               &  90.57 (5.02) &         - (-) &       91.18 (2.16) &      81.20 (3.01) &     72.91 (2.32) \\
all experimental                &  91.28 (4.63) &         - (-) &       88.18 (8.43) &      79.97 (3.16) &     75.98 (2.04) \\
synthetic and good experimental &  97.83 (1.00) &         - (-) &       93.35 (1.20) &      81.52 (3.10) &     72.54 (2.76) \\
synthetic and all experimental  &  98.49 (0.93) &         - (-) &       93.63 (1.20) &      82.66 (2.29) &     77.20 (1.87) \\
synthetic with noise            &  97.52 (1.61) &         - (-) &       88.29 (2.30) &      70.82 (2.29) &     62.59 (2.98) \\
\hline \hline
\end{tabular}
}
 \caption{Accuracies of the convolutional neural network summarised in \autoref{tab:cnn}. Average accuracies and standard deviation are taken over ten train and test splits, which were selected randomly with equal numbers of single and double diagrams.}
    \label{tab:results}
\end{table*}

\begin{table*}[!t]
\resizebox{ \columnwidth}{!}{\begin{tabular}{@{\extracolsep{8pt}}lccccc}
\hline \hline 
  training data & \multicolumn{5}{c}{average accuracy (std)}  \\ \cline{2-6} 

{} &      training &                                       synthetic &                          clean experimental &                               good experimental &                                all experimental \\
\hline
synthetic                       &  99.22 (0.59) &  $\begin{bmatrix}1047&42\\29&1066\end{bmatrix}$ &  $\begin{bmatrix}90&11\\13&88\end{bmatrix}$ &  $\begin{bmatrix}404&144\\166&382\end{bmatrix}$ &  $\begin{bmatrix}728&369\\407&690\end{bmatrix}$ \\
good experimental               &  90.57 (5.02) &                                           - (-) &   $\begin{bmatrix}95&6\\10&91\end{bmatrix}$ &      $\begin{bmatrix}94&15\\23&87\end{bmatrix}$ &    $\begin{bmatrix}181&36\\82&139\end{bmatrix}$ \\
all experimental                &  91.28 (4.63) &                                           - (-) &  $\begin{bmatrix}90&11\\11&90\end{bmatrix}$ &      $\begin{bmatrix}88&21\\23&85\end{bmatrix}$ &    $\begin{bmatrix}169&50\\58&161\end{bmatrix}$ \\
synthetic and good experimental &  97.83 (1.00) &                                           - (-) &   $\begin{bmatrix}98&3\\11&91\end{bmatrix}$ &      $\begin{bmatrix}95&12\\28&83\end{bmatrix}$ &    $\begin{bmatrix}185&33\\89&131\end{bmatrix}$ \\
synthetic and all experimental  &  98.49 (0.93) &                                           - (-) &    $\begin{bmatrix}98&3\\8&93\end{bmatrix}$ &      $\begin{bmatrix}95&15\\24&84\end{bmatrix}$ &    $\begin{bmatrix}180&46\\53&160\end{bmatrix}$ \\
synthetic with noise            &  97.52 (1.61) &                                           - (-) &  $\begin{bmatrix}88&13\\11&91\end{bmatrix}$ &  $\begin{bmatrix}373&175\\145&403\end{bmatrix}$ &  $\begin{bmatrix}601&496\\324&773\end{bmatrix}$ \\
\hline \hline
\end{tabular}
}
 \caption{Confusion matrices of convolutional neural network classification results. Diagonal elements indicate correct classification of single (SD) and double (DD) dots, while off diagonals correspond to false single dot (FSD) and false double dot (FDD): $\begin{bmatrix}SD & FDD\\ FSD & DD\end{bmatrix}$.}
     \label{tab:my_CNN_conf_mat}
\end{table*}

\begin{table*}[!t]
\resizebox{ \columnwidth}{!}{\begin{tabular}{@{\extracolsep{8pt}}lccccc}
\hline \hline 
  training data & \multicolumn{5}{c}{average accuracy (std)}  \\ \cline{2-6} 

{} &      training &     synthetic & clean experimental & good experimental & all experimental \\
\hline
synthetic                       &  99.64 (0.16) &  95.91 (0.46) &       77.19 (3.49) &      64.43 (1.99) &     61.74 (1.57) \\
good experimental               &  91.30 (4.82) &         - (-) &       82.24 (4.07) &      76.82 (2.16) &     70.38 (2.22) \\
all experimental                &  90.61 (3.24) &         - (-) &       83.39 (4.42) &      75.65 (2.66) &     72.59 (1.40) \\
synthetic and good experimental &  98.29 (1.13) &         - (-) &       88.81 (1.37) &      77.58 (2.50) &     69.87 (1.60) \\
synthetic and all experimental  &  96.99 (1.61) &         - (-) &       88.52 (1.75) &      78.36 (2.67) &     72.98 (2.22) \\
synthetic with noise            &  98.65 (1.10) &         - (-) &       75.86 (4.33) &      64.23 (3.43) &     61.01 (2.51) \\
\hline \hline
\end{tabular}
}
 \caption{Accuracies of a shallower convolutional neural network, with one convolutional layer and four dense layers. The parameters used are the same as in \autoref{tab:cnn}, but with the first convolutional layer with 32 kernels only. Average accuracies and standard deviation are taken over ten train and test splits, which were selected randomly with equal numbers of single and double diagrams.}
     \label{tab:results_shallow_cnn}
\end{table*}

\begin{table*}[!t]
\resizebox{ \columnwidth}{!}{\begin{tabular}{@{\extracolsep{5pt}}lccccccc}
\hline \hline 
   training \&  transfer data &   \multicolumn{7}{c}{average accuracy (std)} \\  \cline{2-8} 
  & transfer training   & \multicolumn{2}{c}{clean experimental} & \multicolumn{2}{c}{good experimental} & \multicolumn{2}{c}{all experimental} \\  \cline{2-2} \cline{3-4} \cline{5-6} \cline{7-8} 

{} &       &  pre-train  &   transfer  &  pre-train  &   transfer  &  pre-train  &   transfer  \\
\hline
synthetic and good experimental       &  70.21 (3.88) &  86.81 (3.48) &  87.52 (3.74) &  71.00 (5.06) &  72.64 (4.96) &  64.45 (4.44) &  66.67 (3.41) \\
synthetic and all experimental        &  64.59 (3.24) &  86.64 (3.53) &  86.36 (3.92) &  71.51 (5.21) &  72.91 (4.64) &  64.06 (3.80) &  66.46 (3.53) \\
noisy synthetic and good experimental &  70.25 (2.55) &  88.31 (2.49) &  89.44 (2.28) &  69.84 (3.51) &  72.61 (3.26) &  63.18 (2.79) &  64.04 (2.90) \\
noisy synthetic and all experimental  &  62.28 (3.42) &  88.28 (2.41) &  89.13 (2.32) &  69.89 (3.59) &  72.92 (3.47) &  61.65 (4.15) &  64.27 (3.16) \\
\hline \hline
\end{tabular}
}
 \caption{Transfer learning results of convolutional neural network summarised in \autoref{tab:cnn}.  Average accuracies and standard deviation are taken over ten train and test splits, which were selected randomly with equal numbers of single and double diagrams.}
     \label{tab:transfer_results}
\end{table*}

\begin{table*}[!t]
\resizebox{ 0.9\columnwidth}{!}{\begin{tabular}{@{\extracolsep{8pt}}lcccc} 
 \hline \hline 
 & \multicolumn{4}{c}{average accuracy (std)}  \\ \cline{2-5} 
max noise amplitude  & training & clean experimental & good experimental & all experimental \\
\hline \multicolumn{5}{c}{No noise} \\ \cline{1-5} {}

0 &  99.22 (0.59) &       87.15 (3.15) &      72.30 (3.40) &     64.62 (2.25) \\

\hline  \hline 
\multicolumn{5}{c}{pinchoff modulation} \\ \cline{1-5} {}

0.05 &  97.71 (1.10) &       86.58 (4.00) &      70.85 (3.73) &     63.56 (4.11) \\
0.1  &  97.20 (1.49) &       83.43 (7.71) &      67.65 (4.88) &     60.64 (3.94) \\
0.2  &  96.56 (1.27) &       83.96 (6.88) &      66.84 (4.69) &     59.81 (3.69) \\
0.3  &  97.81 (1.16) &       84.46 (5.75) &      67.97 (4.06) &     60.48 (4.30) \\

\hline  \hline 
\multicolumn{5}{c}{random telegraph} \\ \cline{1-5} {}

0.05 &  98.22 (1.10) &       86.76 (4.25) &      70.61 (2.83) &     62.76 (2.41) \\
0.1  &  97.10 (1.46) &       84.25 (6.91) &      70.24 (5.28) &     62.98 (4.55) \\
0.2  &  96.45 (1.34) &       82.88 (5.07) &      67.21 (5.63) &     59.79 (4.59) \\
0.3  &  97.72 (0.80) &       89.49 (3.50) &      71.49 (3.20) &     63.68 (2.70) \\

\hline  \hline 
\multicolumn{5}{c}{charge jumps} \\ \cline{1-5} {}

0.05 &  98.03 (1.00) &       87.21 (2.68) &      70.39 (3.00) &     62.90 (3.06) \\
0.1  &  97.39 (1.58) &       85.37 (6.52) &      69.17 (4.95) &     61.94 (3.97) \\
0.2  &  96.89 (1.57) &       83.01 (8.38) &      66.97 (5.78) &     61.56 (4.13) \\
0.3  &  97.10 (0.47) &       86.70 (3.08) &      72.11 (2.96) &     64.00 (2.03) \\

\hline  \hline 
\multicolumn{5}{c}{1/f} \\ \cline{1-5} {}

0.05 &  97.75 (1.13) &       88.34 (3.38) &      71.08 (3.32) &     63.12 (3.57) \\
0.1  &  97.94 (1.10) &       87.33 (3.62) &      70.02 (2.95) &     62.67 (3.79) \\
0.2  &  96.52 (1.34) &       81.98 (5.77) &      67.74 (3.67) &     61.03 (4.51) \\
0.3  &  97.22 (1.18) &       86.49 (3.76) &      70.04 (2.64) &     62.04 (2.97) \\

\hline  \hline 
\multicolumn{5}{c}{random current modulation} \\ \cline{1-5} {}

0.05 &  97.92 (1.12) &       87.77 (3.83) &      69.95 (3.74) &     62.38 (3.44) \\
0.1  &  97.48 (1.07) &       85.56 (4.77) &      69.61 (4.33) &     63.06 (4.28) \\
0.2  &  96.84 (1.68) &       84.23 (5.91) &      68.27 (3.67) &     61.61 (3.43) \\
0.3  &  97.21 (2.25) &       83.22 (5.92) &      67.74 (5.38) &     62.18 (3.33) \\

\hline  \hline 
\multicolumn{5}{c}{white} \\ \cline{1-5} {}

0.05 &  98.15 (1.03) &       88.70 (3.38) &      70.93 (3.08) &     63.88 (3.11) \\
0.1  &  96.73 (1.45) &       84.07 (7.13) &      69.70 (5.20) &     63.21 (4.34) \\
0.2  &  96.97 (1.60) &       83.24 (7.27) &      68.17 (4.49) &     61.75 (4.00) \\
0.3  &  96.47 (1.44) &       81.56 (7.72) &      65.74 (4.25) &     59.07 (3.25) \\

\hline  \hline 
\hline \hline
 \end{tabular}
}
 \caption{Noise model study showing classification accuracies of the convolutional neural network summarised in \autoref{tab:cnn}.  Each noise type is added individually and at varying amplitudes. Average accuracies and standard deviation are taken over ten train and test splits, which were selected randomly with equal numbers of single and double diagrams.}
     \label{tab:noise_results}
\end{table*}

\begin{table*}[!t]
 \resizebox{ \columnwidth}{!}{\begin{tabular}{@{\extracolsep{8pt}}lccccc}
\hline \hline 
 training data & \multicolumn{5}{c}{average accuracy (std)}  \\ \hline 
\multicolumn{6}{c}{LogisticRegression}  \\ \cline{1-6} 
 
{} &      training &     synthetic & clean experimental & good experimental & all experimental \\
\hline
synthetic                       &  99.61 (0.10) &  87.43 (0.66) &       46.57 (3.69) &      47.53 (2.65) &     49.07 (1.13) \\
good experimental               &  99.94 (0.06) &         - (-) &       68.48 (2.85) &      78.86 (1.81) &     69.39 (2.48) \\
all experimental                &  99.63 (0.16) &         - (-) &       69.04 (3.50) &      75.83 (2.84) &     71.52 (2.28) \\
synthetic and good experimental &  98.12 (0.12) &         - (-) &       59.57 (3.10) &      67.70 (3.54) &     62.58 (2.71) \\
synthetic and all experimental  &  97.27 (0.11) &         - (-) &       61.79 (4.77) &      68.71 (3.33) &     65.23 (2.11) \\
synthetic with noise            &  94.82 (0.19) &         - (-) &       48.30 (3.39) &      54.21 (2.02) &     53.53 (1.50) \\
\hline \hline\multicolumn{6}{c}{MLPClassifier}  \\ \cline{1-6} 
 
{} &       training &     synthetic & clean experimental & good experimental & all experimental \\
\hline
synthetic                       &   99.57 (0.44) &  94.09 (0.57) &       50.33 (2.75) &      49.53 (1.47) &     49.98 (1.14) \\
good experimental               &   96.92 (1.94) &         - (-) &       77.47 (4.28) &      81.13 (3.43) &     70.67 (3.09) \\
all experimental                &   95.41 (2.26) &         - (-) &       75.44 (6.67) &      80.25 (3.01) &     75.51 (2.51) \\
synthetic and good experimental &   99.46 (0.81) &         - (-) &       65.43 (6.15) &      80.12 (3.18) &     71.09 (2.39) \\
synthetic and all experimental  &   99.02 (0.46) &         - (-) &       65.90 (5.70) &      81.15 (2.64) &     75.49 (2.86) \\
synthetic with noise            &  100.00 (0.00) &         - (-) &       54.83 (2.95) &      54.61 (1.62) &     55.03 (0.92) \\
\hline \hline\multicolumn{6}{c}{DecisionTreeClassifier}  \\ \cline{1-6} 
 
{} &      training &     synthetic & clean experimental & good experimental & all experimental \\
\hline
synthetic                       &  92.52 (0.33) &  90.47 (0.82) &       54.70 (4.79) &      50.77 (2.84) &     51.18 (1.93) \\
good experimental               &  83.39 (2.75) &         - (-) &       60.08 (7.25) &      67.10 (3.43) &     62.91 (2.97) \\
all experimental                &  77.92 (2.01) &         - (-) &       57.54 (6.74) &      68.91 (3.86) &     64.55 (2.65) \\
synthetic and good experimental &  90.09 (0.41) &         - (-) &       66.12 (1.36) &      61.61 (4.20) &     59.79 (2.52) \\
synthetic and all experimental  &  88.12 (0.44) &         - (-) &       65.66 (2.68) &      62.66 (3.09) &     60.31 (2.84) \\
synthetic with noise            &  90.74 (0.35) &         - (-) &       51.23 (0.84) &      50.43 (0.33) &     48.54 (0.62) \\
\hline \hline\multicolumn{6}{c}{RandomForestClassifier}  \\ \cline{1-6} 
 
{} &      training &     synthetic & clean experimental & good experimental & all experimental \\
\hline
synthetic                       &  86.54 (0.76) &  85.53 (0.70) &       48.73 (2.67) &      50.43 (0.98) &     50.13 (0.64) \\
good experimental               &  79.87 (1.99) &         - (-) &       65.81 (4.56) &      65.48 (3.84) &     61.49 (2.00) \\
all experimental                &  72.61 (1.56) &         - (-) &       63.47 (3.69) &      61.34 (4.06) &     63.29 (3.04) \\
synthetic and good experimental &  83.28 (0.70) &         - (-) &       48.33 (2.26) &      49.24 (3.02) &     49.88 (2.05) \\
synthetic and all experimental  &  80.80 (0.63) &         - (-) &       48.91 (2.00) &      49.72 (2.95) &     49.93 (2.35) \\
synthetic with noise            &  83.91 (0.54) &         - (-) &       50.00 (0.40) &      49.82 (0.18) &     49.84 (0.08) \\
\hline \hline\multicolumn{6}{c}{KNeighborsClassifier}  \\ \cline{1-6} 
 
{} &      training &     synthetic & clean experimental & good experimental & all experimental \\
\hline
synthetic                       &  91.46 (0.34) &  84.76 (0.76) &       47.02 (2.56) &      48.75 (1.49) &     51.82 (0.95) \\
good experimental               &  88.74 (0.89) &         - (-) &       75.14 (3.08) &      75.16 (3.02) &     67.37 (2.78) \\
all experimental                &  87.39 (0.90) &         - (-) &       73.44 (3.02) &      76.56 (3.55) &     72.10 (2.64) \\
synthetic and good experimental &  91.42 (0.19) &         - (-) &       60.97 (3.32) &      75.87 (3.24) &     67.19 (2.43) \\
synthetic and all experimental  &  90.72 (0.19) &         - (-) &       66.44 (3.41) &      74.37 (2.85) &     69.80 (2.78) \\
synthetic with noise            &  90.06 (0.71) &         - (-) &       57.67 (4.41) &      54.29 (2.53) &     54.31 (1.63) \\
\hline \hline\multicolumn{6}{c}{SVC}  \\ \cline{1-6} 
 
{} &       training &     synthetic & clean experimental & good experimental & all experimental \\
\hline
synthetic                       &   99.95 (0.02) &  85.57 (0.71) &       48.37 (1.22) &      48.13 (2.09) &     49.02 (0.79) \\
good experimental               &  100.00 (0.00) &         - (-) &       69.67 (3.48) &      77.92 (3.29) &     68.90 (2.42) \\
all experimental                &   99.91 (0.06) &         - (-) &       67.91 (4.58) &      74.62 (4.25) &     68.17 (2.23) \\
synthetic and good experimental &   99.03 (0.15) &         - (-) &       57.32 (3.77) &      66.31 (2.01) &     61.63 (2.33) \\
synthetic and all experimental  &   98.40 (0.08) &         - (-) &       59.80 (4.19) &      68.35 (3.85) &     65.14 (2.41) \\
synthetic with noise            &   96.10 (0.13) &         - (-) &       47.81 (3.05) &      53.49 (1.60) &     51.48 (1.65) \\
\hline \hline
\hline \hline
\end{tabular}
}
 \caption{Parametric binary classification results. Average accuracies and standard deviation are taken over ten train and test splits, which were selected randomly with equal numbers of single and double diagrams.}
    \label{tab:binary_results}
\end{table*}

\end{document}